\documentclass[12pt,preprint]{emulateapj}
\slugcomment{ApJ, in press}

\shorttitle{CSM Bubble}
\shortauthors{Laming \& Hwang}

\begin{document}

\title{The Circumstellar Medium of Cassiopeia A Inferred from the Outer Ejecta
Knot Properties}


\author{Una Hwang\altaffilmark{1} \& J. Martin Laming\altaffilmark{2}}


\altaffiltext{1}{NASA/GSFC Code 662, Greenbelt MD 20771, and the Johns Hopkins University, Baltimore MD 21218
\email{Una.Hwang-1@nasa.gov}}
\altaffiltext{2}{Space Science Division, Naval
Research Laboratory, Code 7674L, Washington DC 20375
}

\begin{abstract}

We investigate the effect of the circumstellar medium density profile
on the X-ray emission from outer ejecta knots in the Cassiopeia A
supernova remnant using the 1 Ms Chandra observation. The spectra of a
number of radial series of ejecta knots at various positions around
the remnant are analyzed using techniques similar to those devised in
previous papers. We can obtain a reasonable match to our data for a
circumstellar density profile proportional to $r^{-2}$ as would arise
from the steady dense wind of a red supergiant, but the agreement is
improved if we introduce a small (0.2-0.3 pc) central cavity around
the progenitor into our models. Such a profile might arise if the
progenitor emitted a fast tenuous stellar wind for a short period
immediately prior to explosion. We review other lines of evidence
supporting this conclusion. The spectra also indicate the widespread
presence of Fe-enriched plasma that was presumably formed by complete
Si burning during the explosion, possibly via alpha-rich
freezeout. This component is typically associated with hotter and more
highly ionized gas than the bulk of the O- and Si-rich ejecta.

\end{abstract}

\section{Introduction}

More than three hundred years after the event, the optical supernova
spectrum of the explosion that formed the Cassiopeia A supernova
remnant (SNR) has been observed in a scattered light echo and
establishes that the supernova was of type IIb (Krause et al. 2008).
The spectrum strongly resembles that of the IIb prototype SN 1993J at
maximum, with broad H$\alpha$ and weak He lines indicating that the
red supergiant exploded with a thin layer of the H envelope
remaining outside the He core.  The IIb supernovae form a transition
between the Type II explosions of massive stars that retain a
substantial H layer and the Type Ib explosions of stars that have
entirely lost their H envelope in a presupernova stellar wind; the
late-time spectra of IIb events transform to resemble those of Ib's.
For Cas A, the IIb classification is consistent with the presence of a
number of fast ejecta knots showing optical H emission
\citep[e.g., as discussed by][]{chevalier03}.

Cas A's echo spectrum is distinguished from those of normal Type II
supernovae by the presence of broad absorption lines in the spectrum
due to the Doppler broadening of the rapidly expanding ejecta and the
absence of prominent unresolved lines.  This led \citet{krause08} to
suggest that dense circumstellar material did not reach to the surface
of the Cas A progenitor, but rather that the progenitor created a
small bubble in the circumstellar medium by emitting a fast stellar
wind prior to explosion. Indeed, WNL stars that have H levels
comparable to those seen in the fast H knots in Cas A are capable of
generating fast winds to blow such a bubble (Lamers \& Nugis 2002).
The actual extent of such a wind is constrained by the presence of
prominent ``jet'' structures in Cas A today, which are now
demonstrated to be associated with asymmetries in the explosion
(Laming et al. 2006). In the most extreme cases, emerging jets could
be destroyed by an encounter with a swept-up mass shell on the
periphery of a bubble.  To avoid destruction of the jets, the duration
of a pre-explosion fast wind in the Cas A progenitor is limited by
hydrodynamical simulations to less than about 2000-2500 years
\citep{schure08}.  This places the maximum radius for a bubble at
0.3-.35 pc.  Hydrodynamic considerations lead \citet{vanveelen09} to
similarly conclude that the Cas A progenitor cannot have undergone a
substantial WR phase.

There have been a number of previous arguments favoring the presence
of a circumstellar bubble.  A short ($\sim 10^4$ yr) Wolf-Rayet phase
has been suggested for Cas A by \citet{garcia96} to provide
consistency for the results of their simulations of the circumstellar
structure with the low velocities and small radii measured for the
optically emitting quasi-stationary flocculi (QSFs) associated with
the CSM.  In the work of \citet{laming03}, radial series of X-ray
emitting ejecta knots in Cas A were examined in the context of the
one-dimensional hydrodynamics of ejecta, assuming an ejecta density
profile with a power-law envelope and a constant density core evolving
into a $\rho \propto r^{-2}$ circumstellar environment.  The
inconsistency in this model of blast wave velocities slightly higher
and radii lower than those measured directly by \citet{delaney03}, can
also be resolved if the remnant evolved for a time in a small
circumstellar bubble such as might be formed during a short Wolf-Rayet
phase.  (Alternatively, \citet{patnaude09} suggest that the observed
forward shock velocity and radii can be matched instead by the
diversion of energy into the acceleration of cosmic rays.)

In this paper we explore these ideas further using the 1 Ms X-ray
observation of Cas~A obtained by Chandra \citep{hwang04}.  As in
previous work by \citet{laming03}, \citet{hwang03} and
\citet{laming06}, we interpret the ejecta knots and filaments that
dominate the X-ray appearance of Cas A as ``knots'' of distinctive
ejecta composition rather than as knots of substantially higher
density, with the rationale that highly overdense knots will be efficiently
disrupted by hydrodynamical instabilities. Here we will examine the
spectra of several radial series of ejecta knots in the context of
hydrodynamical models that incorporate a small circumstellar bubble
within the red supergiant wind.

\section{X-ray Spectral Analysis}
\subsection{Introduction}
For the X-ray spectral study, we use a 1 Ms observation with the
backside-illuminated S3 chip of the Advanced CCD Imaging Spectrometer
(ACIS) on the Chandra Observatory (Hwang et al. 2004).  These data
were obtained in 2004 in nine observation segments, all but one of
which occurred over a period of three consecutive weeks.  The data
extraction procedure essentially follows that described in
\citet{laming06}, with updated calibrations (CIAO-4.0, CALDB-3.4)
applied to the data.  The data were cleaned to exclude non-standard
event grades, bad pixels, and time intervals with high background
levels to yield 980 ks of exposure time.  The data were also corrected
for the time-dependent gain of the detector, but could not be
corrected for charge-transfer inefficiency given that the events were
characterized on board the satellite before telemetry (i.e., the data
were obtained in Chandra GRADED mode).

\begin{figure}
\plottwo{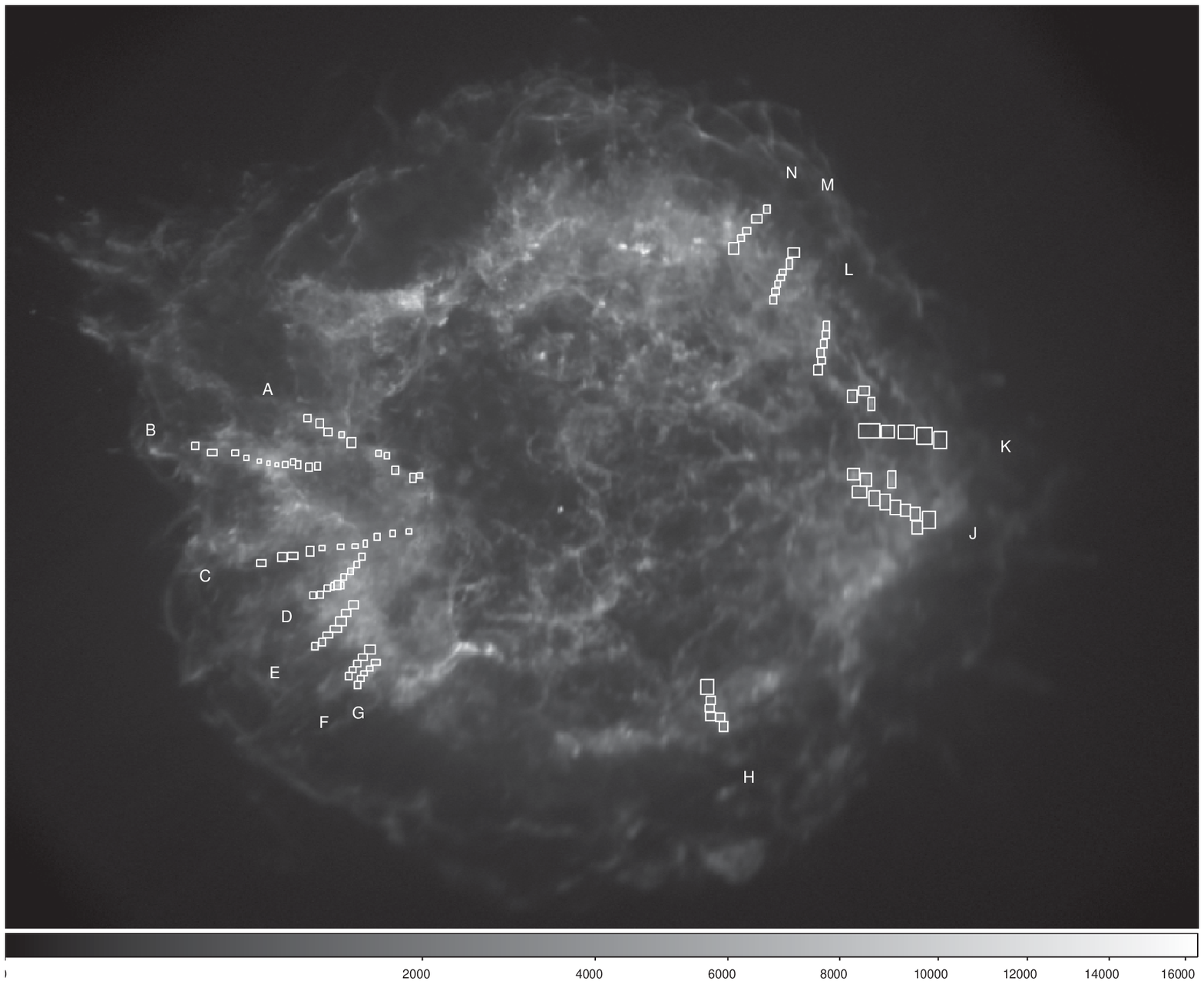}{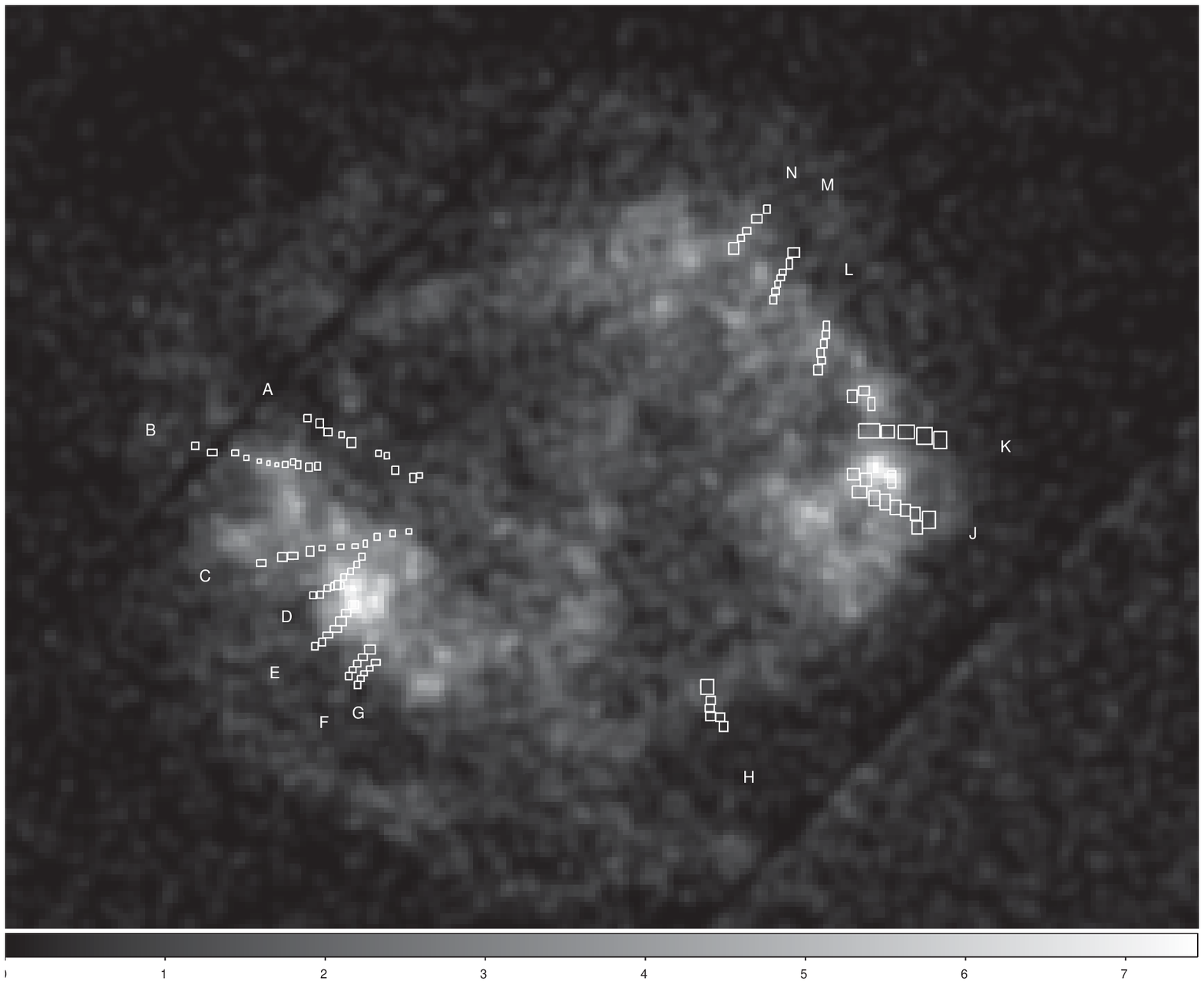}

\caption{(Left:) Radial series of knots labelled by letter overlaid on
  the Chandra ACIS broadband image of Cas A.  The regions are numbered
  from the inside out towards the remnant edge.
(Right:) The same regions overlaid on the smoothed 8-15 keV XMM-Newton PN image from 2000.
}
\label{knots}
\end{figure}

A number of approximately radial ``series'' of several knots each were
chosen to sample the spectra at a range of azimuthal locations as
shown in Figure \ref{knots}.  The size of the spectral regions varied
from 1.5$''$ to 6$''$ on a side, with most averaging 2-3$''$.  The
spectra are of very high quality, typically containing from a few to
several tens of thousands of counts.  A single spectrum and
corresponding spectral calibration files were calculated for each
region by weighting the individual contributions from each observation
segment according to the relative exposure time, and then combining
them.  The spectral background was taken from a single set of
off-source regions on the detector.

We consider ejecta regions only.  We then assume that the continuum
comes mainly from ionized O and heavier elements, and accordingly set
the abundances of the lighter elements to zero.  It is also possible
to model the spectra with a light element continuum including H and
He.  The assumption of which light elements provide the continuum does
affect inferences about electron densities and masses in the knot.
While some H and He is certainly present in the remnant given its Type
IIb classification, the Cas A progenitor exploded at only 4 M$_\sun$
\citep{young06}, so it should have lost most of the gas in its outer
layers.  Moreover, the regions we select are taken to be enriched in
heavy elements by instabilities during the explosion.

The basic spectral model we use is a single-temperature nonequilibrium
ionization (NEI) plasma modified by interstellar absorption.  This model has
a single average ionization age (defined by the product of electron
density and the time since shock passage).  Given the known,
significant bulk velocities of the ejecta-dominated gas
\citep[e.g.,][]{markert83}, we also allow for additional Gaussian
broadening above the detector response, and a red- or blue-shift for
the plasma.  We thereby obtain the average electron temperature and
ionization age, the element abundances, and a measure of the
line-of-sight velocity.

Spectral complexity is seen, however, at a number of levels.  First,
plane-parallel shock models, which include a range of ionization ages
from zero up to a fitted maximum, were generally more successful with
our data than NEI models, mainly because they are better able to
characterize the shape of the Fe L blend.  While plane-parallel shock
spectral models do not increase the number of fitted parameters, they
are more difficult to interpret in the context of the hydrodynamical
models discussed in section 3.  For that reason, we favor the NEI
models for this discussion.  The actual fitted parameters are quite
comparable, in the sense that the temperatures are similar and the
maximum ionization age in the plane-parallel shock model generally
turns out to be about twice the average ionization age in the NEI
models used here.

\begin{figure}
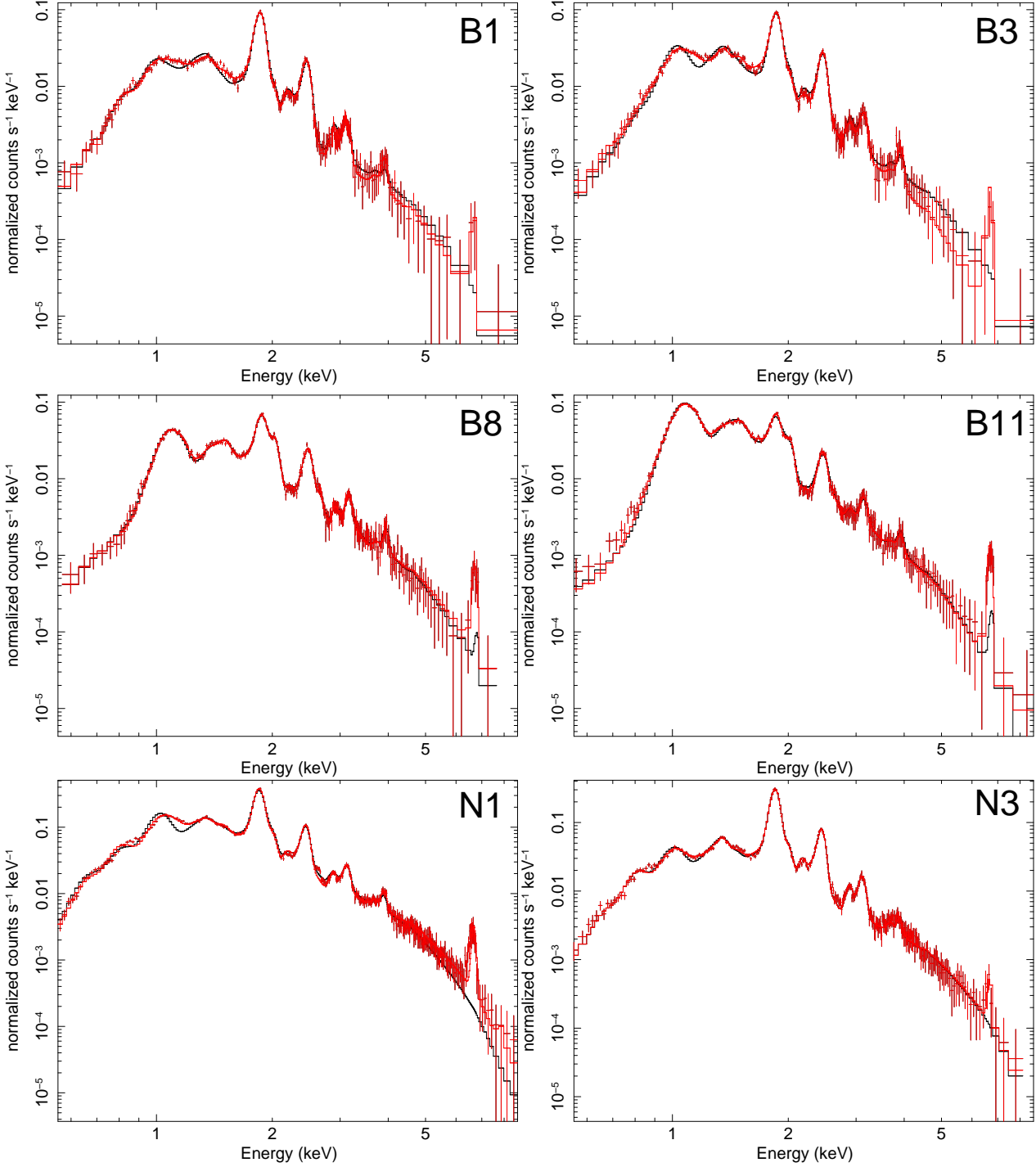

\centerline{\includegraphics[scale=0.35,angle=-90]{f2a.ps}\includegraphics[scale=0.35,angle=-90]{f2b.ps}}
\centerline{\includegraphics[scale=0.35,angle=-90]{f2c.ps}\includegraphics[scale=0.35,angle=-90]{f2d.ps}}
\centerline{\includegraphics[scale=0.35,angle=-90]{f2e.ps}\includegraphics[scale=0.35,angle=-90]{f2f.ps}}
\caption{Selected spectra from the B and N series showing both single (black) and two (red) NEI models as described in the text.}
\label{spectra}
\end{figure}

Second, most of the spectra show a Fe K blend which is left
unaccounted for by the single-component NEI models.  Even
single-component plane-parallel shock models are successful in this
respect in only a very limited number of cases, typically those where
the Fe K blend is especially prominent.  Such spectra usually also
feature a prominent Fe L blend, and high ratios of the Si Ly $\alpha$
to He $\alpha$ emission (for example, see knots B8 and B11 in Figure
\ref{spectra}).  In the remaining majority of cases where single
component NEI and plane-parallel shock models are both inadequate, the
addition of a second spectral component including {\it only} Fe and Ni
does provide significantly improved fits.

Our final spectral model is thus a two-component NEI model, with one
component corresponding to O-rich ejecta including all the abundant
elements heavier than O, and the second including only Fe and Ni.  The
redshift of the Fe component was fixed at 0 in order to better
constrain the other parameters.  In most cases, this second component
did act to fill in the Fe K blend.  In just a few instances, it
preferred instead to improve the shape of the Fe L blend; those cases,
however, might be better handled with a range of ionization ages.
Throughout, the O-rich ejecta component remains dominant, and its
properties are generally not much changed from the single NEI case,
though there are a few exceptions.  Generally, these reduce the high
temperatures of the main ejecta component for some of the knots
(mostly in the B and E series), although there were also a few
instances where the ejecta temperature increased.  We discuss the Fe
components a little further in the following subsection.

The fitted temperatures for the main ejecta component for these knots
are typically 1-2 keV, and the average ionization ages are most often
in the low- to mid- 10$^{11}$ cm$^{-3}$s range.  Significant
differences are seen amongst the various sets, with somewhat higher
temperatures measured for certain knots, especially in series A, D, J,
L, M, and N.  The knots in the western region of the remnant tend to
have higher fitted temperatures on the whole.  The column density
varies significantly across the remnant, as has been frequently noted
in previous observations (e.g., Eriksen et al. 2009, Keohane et
al. 1996, Troland et al. 1985), with the highest values in this set
seen in the west (i.e., series J and K), where a molecular cloud
is known to be interacting with and obscuring the remnant.

We give all the two component NEI fit results in Table 1, but rather
than showing the spectra for all hundred or so knots considered here,
we do so only for selected knots from two particular radial series (B
and N) in Figure \ref{spectra}.  The knots in series B generally have
a strong Fe K blend and show a strong evolution in ionization age that
can be traced by the Si Ly $\alpha$/He $\alpha$ ratio.  The knots in
series N are representative of those that have weaker Fe K emission
and more uniform spectral properties.  For these and all other sets of
knots, we plot the fitted temperatures and ionization ages of the
knots in Figure \ref{tnet}, together with the models that will
be described shortly in section 3.

\subsection{Fe plasma component}

As noted above, no single-component NEI models were able to reproduce
the entire spectrum of any knot showing an Fe K blend, whereas two
such components did in most cases provide significantly improved fits
to the spectrum, including the Fe K emission.  While the Fe
component is not the focus of this study and we do not undertake a
detailed interpretation of it here, we do note that the fitted
temperatures of the Fe component are generally rather high at above 2 keV,
and that ionization ages are often, though not always, higher than
the main ejecta component--generally a few to several $10^{11}$
cm$^{-3}$s.

Aside from the obvious indication of widespread spectral complexity
even on 2-3$''$ angular scales, it appears that the spectra of many of
the knots do require an additional emission component that is
consistent with pure Fe or very high enrichment in Fe.  In their early
Chandra assay of ejecta knots in Cas A, Hughes et al. (2000) note that
strong Fe K emission accompanying strong Si emission likely indicates
the presence of additional Fe ejecta beyond the relatively low Fe yield
produced by incomplete Si burning.  Presumably, this additional Fe is
nearly pure Fe from complete Si burning with or without alpha-rich
freezeout.  The results of \citet{willingale02} suggest that certain
lines of sight should indeed include a juxtaposition of ejecta of
various compositions and velocities.  Our results support the
idea that an Fe-enriched ejecta component is widespread throughout the
remnant in the regions that we sampled.

Second, this Fe enriched plasma tends to be hotter and is probably
more highly ionized than the main ejecta component that includes O,
Si, and other elements.  In the models that we will describe in the
next section, these ionization ages correspond to an ejecta Lagrange
mass coordinate of $\sim 0.4$, where 0 corresponds to the outermost
layers of ejecta and 1 to the mass cut at the presumed neutron
star. Thus for 2 $M_{\sun}$ ejecta mass, the Fe is at a significant
distance of approximately 1.2 $M_{\sun}$ from the mass cut.  In the
cases where highly enriched Fe plasma seems to have been well-mixed
with other plasma (i.e., those knots where a simple plane-parallel
shock component could describe all the emission), there appears to be
a tendency for even higher degrees of ionization.  Marked evolution in
ionization age can be seen in the radial series located in the eastern
region of the remnant, e.g., series B, C, D, and E. Such high
ionization ages are not commonly seen in Cas A, and as it turns out,
the eastern region in the remnant is where a high degree of Fe
enrichment is most clearly observed \citep{hwang03}.

We defer a comprehensive study of the properties of this Fe emission
for a forthcoming paper.  For the time being, we restrict ourselves to
the observation that these spectra clearly show a widespread
contribution from plasma enriched in Fe that have distinct plasma
properties from the other ejecta plasma that dominates the emission.

\subsection{Forward shock and nonthermal emission}
Aside from the spectral complexity due to multiple ejecta components,
it is also possible that there will be some significant emission
projected from the forward shock, even in these ejecta-dominated
regions.  Recent work has shown that the forward shock emission in
young remnants is more strongly dominated by nonthermal than thermal
processes.  Moreover, in Cas A, even the interior regions of the
remnant have been shown to be associated with hard nonthermal X-ray
emission \citep{bleeker01, helder08}.  We assessed the location of our
ejecta knots on the 8-15 keV hard continuum image obtained from a 2000
observation with the PN detector on XMM-Newton (obsid 0097610801,
which is shown here in Figure 1, and is similar to that in
Bleeker et al. 2001), as well as the 4-6 keV Chandra continuum image.
Contamination from nonthermal emission is more likely in the west,
where it is strongest, but the Chandra images show that the hard
emission is distributed in highly filamentary features throughout the
remnant. Precise coincidences between our chosen region and the hard
X-ray emission are not common, but given that the 4-6 keV emission is
variable with a timescale of order a year \citep{uchiyama08,
  patnaude07, patnaude09}, one might expect the 8-15 keV continuum to be similarly
variable if it is produced by the same mechanism.

All of the knots presented here are true ejecta-dominated features,
but the presence of a nonthermal component could in principle
significantly affect the inferred temperatures and abundances for the
ejecta knots.  The addition of a power-law component certainly does
change the fitted temperature of the thermal ejecta component in some
knots with particularly strong continuum that we have not included
here.  Their initially high fitted temperatures of 2-3 keV were
reduced significantly with the introduction of a power-law continuum
to the model, while the uncertainties in the temperatures increased
significantly.  In virtually all the cases considered here, however,
the line emission from the ejecta is strongly dominant so that the
fits are driven more strongly by the ejecta line emission than by the
high-energy continuum.  In the few cases where there is a clear
deficit in the continuum at high energies for single component NEI
models---for example in knots H5 and H6---we generally found that
fitting a power-law component of the expected $\sim$2.7 slope (rather
than a second NEI component) did little to change the fitted
temperature for the main ejecta component.
The fact that higher plasma temperatures are observed in region in the
west where the hard emission is stronger is very interesting, however,
in that it may have implications for the nature of this emission, as
regards to whether it is produced by synchrotron emission or
nonthermal bremsstrahlung.  We discuss this a little further in
section 4.2.

\medskip
\section{Modeling and Interpretation}

We model the spectra of the outer knots of Cas A using the
approximations and methods developed in \citet{laming03} and
\citet{hwang03}, extended to deal with the case of expansion into a
stellar wind incorporating a ``bubble'' around the supernova. This is
presumed to arise from a short spell of fast tenuous stellar wind just
prior to the supernova explosion. In our treatment, a zero density
spherical bubble is centered on the supernova, and the density obeys
$\rho\propto 1/r^s$ outside the bubble. We do not include the effect
of a swept-up shell of circumstellar material at the bubble periphery,
but merely consider a pure stellar wind density profile with a cavity
in the center.

We take the model in \citet{laming03} for a $1/r^s$ (with $s=2$)
circumstellar density profile with no bubble as a ``fiducial'' model,
and approximate that the forward shock velocity in the model
including the bubble should be the same as that in the fiducial
model when the same mass of circumstellar plasma has  been swept up.
Then the forward shock radii in the fiducial
case, $r_{b0}$, and in the real case, $r_b$, are related by
\begin{equation}
r_{b0}^3=r_b^3-r_b^sr_{bub}^{3-s}
\end{equation}
where $r_{bub}$ is the bubble radius. (We reference all quantities for
the fiducial model with a 0 subscript.) The ejecta are treated as
having a constant density core, surrounded by an outer envelope with
density $\rho\propto 1/r^n$. Throughout this work we take $n=10$,
following the treatment of \citet{matzner99} for blast wave
propagation through the outer layers of a highly stripped
progenitor. Solutions for forward and reverse shock motion in the core
and envelope phases (i.e. when the reverse shock is propagating
through the ejecta core or envelope) can be found
separately and coupled at the transition time. For $s>0$,
\citet{laming03} adopted a simpler approximation of extending the
envelope blast wave trajectory into the core phase, and coupling it
directly to the appropriate form of the blast wave in the Sedov-Taylor
limit, which occurs at a time $t_{conn}$ \citep[equations A10 - A12
  in][]{laming03}. We adapt this procedure to the case with a bubble
as follows.

While the ejecta envelope is still interacting with the bubble wall,
the forward shock velocity is
\begin{equation}
v_b\left(t\right)=\left({n-3\over n-s}\right)\left({r_b-r_{bub}\over
t-t_{bub}}\right)=v_{b0}\left(t\right)=\left({n-3\over
n-s}\right){r_{b0}\over t_0},
\end{equation}
giving
\begin{equation}
t={t_0\over r_{b0}}\left(r_b-r_{bub}\right) +t_{bub}
\end{equation}
where $t_{bub}$ is taken to be the time at which the forward shock
that is driven into the stellar wind exterior to the bubble has been
accelerated to $v_{core}$, which is the expansion velocity of the
ejecta at the core-envelope boundary. This time is calculated from
equation 3.20 for the forward shock radius in \citet{chevalier89}, by
taking the time derivative and equating it to $v_{core} =
\sqrt{10\left(n-5\right)/3/\left(n-3\right)}$ (in units of
$7090\sqrt{E_{51}/M_{ej}}$ km s$^{-1}$ where $E_{51}$ is the explosion
energy in $10^{51}$ ergs and $M_{ej}$ is the ejecta mass in solar
masses) to give
\begin{equation}
t_{bub}=\left[2\rho
v_b^2r_{bub}^{n-2}\left(n-4\right)\over\left(n-3\right)KA\right]^{1/\left(n-5\right)}.
\end{equation}
Here
$K=0.8\left[1+1.25/\left(n-5\right)\right]^{\left(n-2\right)/3}$ for
$\gamma = 5/3$ gas, and $A$ is defined by the ejecta envelope
density $\rho _e=Ar^{-n}t^{n-3}$ to give
$A=\left(3/4\pi\right)M_{ej}v_{core}^{n-3}\left(n-3\right)/n$.

The initial positions of the contact discontinuity and reverse shock
in the bubble model, $r_{cd}$ and $r_r$ respectively, are estimated as
follows. When the same mass of circumstellar material has been swept
up in both the fiducial and bubble models, the ratio of the densities
of shocked circumstellar gas is
\begin{eqnarray}
{\rho _0\over\rho}&=&{\int _0^{r_{b}}\left(n_0r_{b0}^s/r^s\right)4\pi
r^2dr/\left(4\pi r_{b0}^3/3\right)\over\int
_{r_{bub}}^{r_{b}}\left(n_0r_{b0}^s/r^s\right)4\pi r^2dr/\left(4\pi
\left(r_{b0}^3-r_{bub}^3\right)/3\right)}\nonumber \\ &=&{r_b^3-r_{bub}^3\over
r_b^3-r_b^sr_{bub}^{3-s}}.
\end{eqnarray}
Equating this to
$\left(r_b^3-r_{cd}^3\right)/\left(r_{b0}^3-r_{cd0}^3\right)$ yields
\begin{equation}
r_{cd}=\left[r_b^3-\left(r_{b0}^3-r_{cd0}^3\right){r_b^3-r_{bub}^3\over
r_b^3-r_b^sr_{bub}^{3-s}}\right]^{1/3}.
\end{equation}
The reverse shock position during the envelope phase is given by
\begin{equation}
r_r={r_b-r_{bub}\over l_{ED}} +v_{core}t_{bub}.
\end{equation}
We use this expression in the core phase also, corrected at late
times to ensure that the reverse shock remains behind the contact
discontinuity.

Within this framework of analytic hydrodynamics, we compute the
evolution of the postshock ionization balance and electron and ion
temperatures, using the prescriptions in Appendix B of
\citet{laming03}. A summary of SNR evolution models with various
bubble sizes is given in Table 2. The age of Cas A is constrained by
observations of optical ejecta knots, with the earliest possible
explosion date assuming undecelerated ejecta knots being A.D.
1671.3$\pm 0.9$ \citep{thorstensen01}, and that taking deceleration
into account being A.D. 1681 $\pm 19$ \citep{fesen06}. Given these age
constraints, models with a bubble size of 0.2 - 0.3 pm appear to be
the most likely. Smaller bubbles require explosion dates earlier
than 1671 to allow the blast wave to expand to the observed radius
at its observed velocity near 5000 km s$^{-1}$.  Larger bubbles
correspondingly require later explosion dates, but are
convincingly ruled out by the work of \citet{schure08}.

\section{Results and Discussion}
\subsection{Introduction}
We present models for a variety of elemental abundances as given in
Table 3, which are chosen to illustrate the range of compositions
encountered in our study. While higher concentrations of heavy
elements increase the radiative losses and allow faster cooling, this
is only important once the plasma electron temperature is well below
$10^7$ K, which is outside the range of temperatures that we study
here.  Measured temperatures and ionization ages for knots in all the
radial series are shown in Figure \ref{tnet} against models
including bubbles of radius 0, 0.2, and 0.3 pc calculated with the M4 set of
abundances from Table 3.  The effect of the varying element abundance
on the predicted temperatures and ionization ages is illustrated in
the final panels of the same figure, where we also show the N series
with models for the K10 and C10 set of abundances, which bracket the
range of abundances considered in the models.  The model curves are
seen to be generally similar.

\subsection{Limits on Bubble Size}

Compared to models for evolution into a pure circumstellar wind, the
main important effect of including a bubble is to increase the
temperature of the ejecta knots. This occurs because the knots are
reverse-shocked to approximately the same temperature in both cases,
but at a larger reverse shock radius in the bubble
case. Hence the shocked plasma undergoes less cooling by adiabatic
expansion as it evolves to the current size and age.  This effect is
seen most clearly in the series of knots taken from the
eastern side of the remnant, in Figure \ref{tnet}. Series B, C
and D are seen to favor
bubble models over a pure $1/r^2$ density profile. Series A and F are
ambiguous while series G appears to favor a pure stellar wind
profile.

The presence of a bubble improves agreement between the predicted mass
of radiatively cooled gas and that observed.  The optical emission of
Cas A is completely accounted for by the emission from dense knots of
plasma \citep{hammell08}, leaving no room for emission from plasma
that was initially heated by the shock to X-ray emitting temperatures
and has since cooled by radiation to lower optical (or infra-red)
emitting temperatures. By contrast, \citet{laming03} estimate that
about $0.6 M_{\odot}$ of radiatively cooled gas should be present in
pure O ejecta with an outer ejecta density profile with slope $n=10$
expanding into an unmodified stellar wind.  Pure He ejecta reduces
this radiatively cooled mass to $0.25 M_{\odot}$, but that is still
significantly more mass than is indicated by the
observations. Allowing the remnant to expand into a bubble before
encountering the stellar wind profile reduces the density and hence
the amount of energy lost to radiation. In pure O, the thermal
instability disappears in models with a bubble radius of 0.24 pc, and
in pure He at 0.09 pc.

On the western side of the remnant,
many of the knots (e.g. series L, M, N) show temperatures even higher
than can be accounted for by the bubble models we have considered. As
already noted, the XMM map of emission in the 8-15 keV band \citep[and
  Figure \ref{knots}]{bleeker01} does show high
intensity close to series J, K, L, M, and  N on the west limb, and
close to D and E on the east limb.

The hard X-ray emission in Cas A appears to be largely associated with
ejecta, even though its origin remains under debate, with
\citet{laming01a} and \citet{laming01b} arguing for nonthermal
bremsstrahlung based on X-ray emission over a broad energy range, and
other authors \citep{helder08} concentrating mainly on the 4-6 keV
band and arguing for synchrotron emission.  Nonthermal bremsstrahlung
from nonrelativistic suprathermal electrons as suggested by
\citet{laming01a} and \citet{laming01b} is expected to naturally heat
the ambient thermal electrons by Coulomb collisions. In fact this
Coulomb heating represents a much bigger energy sink for the
suprathermal population than does the radiated bremsstrahlung, as
discussed in detail for the expanding plasma of Cas A by
\citet{laming01b}. By contrast, relativistic electrons emitting
synchrotron radiation do not heat the ambient plasma so
efficiently. In order to have electrons accelerated as cosmic rays at
all, however, electron heating must occur at the shock in order to
provide an injection mechanism. Our models assume no shock electron
heating beyond a simple application of the jump conditions. For
both cases, we might reasonably expect to see higher electron
temperatures than predicted, but the correspondence between
accelerated electrons and plasma heating in the synchrotron case is
less direct than in the nonthermal bremsstrahlung case.



\subsection{Implications for Progenitor}

It is now established that Cas A underwent a Type IIb supernova event
and exploded with at least a thin layer of H intact.  The echo
spectrum presented by \citet{krause08} resembles that of the IIb
prototype SN 1993J, which is inferred to have occured in a binary
system with a massive progenitor of 15-20 M$_\odot$ that evolved to a
3-6 M$_\odot$ He core.  Binary scenarios are also
implicated for other IIb events, such as SN 2008ax \citep{crockett08}.

In terms of enabling a short Wolf-Rayet phase that would create the
required circumstellar bubble, a binary scenario may be more promising
than those for a single star. \citet{woosley93} model a sample of
single massive stars with masses in the range 35 - 85 $M_{\sun}$. Only
stars with initial masses around 60 $M_{\sun}$ are able to expel
enough material to approach $\sim 4 M_{\sun}$ upon explosion, but such
an explosion would not be a Type IIb event, since all the H would have
been lost during the pre-supernova evolution.  \citet{eldridge04},
\citet{perez09}  and
\citet{georgy09} reach similar conclusions.  \citet{young06}
consider in more detail both single and binary progenitor models
specifically for Cas A, with progenitors in the mass range 16-40
$M_{\sun}$. In terms of being able to reproduce the small ejecta
masses at explosion, along with other observables of the Cas A SNR, 16
or 23 $M_{\sun}$ stars with binary companions are strongly
favored. Single stars of 23 or 40 $M_{\sun}$ have final ejecta masses
that are too large, and need explosions that are significantly more
energetic than the accepted $2\times 10^{51}$ ergs if they are to
avoid producing a black hole remnant.  The final parameters inferred
for Cas A by \citet{young06} are similar to those cited above for SN
1993J, at 15-25 M$_\odot$ evolving to 4 M$_\odot$ at explosion.

Binary progenitor systems may indeed be widespread for the
core-collapse supernovae, given that a better match can be obtained to
observed stellar population and core-collapse supernova rates when
binary interactions are taken into account
\citep[e.g.,][]{eldridge08}.  Relevant to the formation of a compact
circumstellar bubble, a binary interaction makes a Wolf-Rayet
phase possible for lower mass stars, and allows the Wolf-Rayet phase
for these stars to be short enough that only a compact bubble would be
formed.

A further constraint is provided by light echo observations in the
infrared.  Aside from the light echo spectrum studied by
\citet{krause08}, light echoes from Cas A have also been observed by
\citet{rest08}, and in the infra-red by \citet{dwek08}. This last
observation does not represent the direct scattering of optical light
from the explosion, but rather the heating of dust by EUV-UV radiation
associated with shock breakout, followed by reradiation at infra-red
wavelengths. The infra-red echoes are located externally to Cas A, and
so the illuminating EUV-UV radiation must have traveled through the
RSG wind from the interior. \citet{dwek08} consider cases of dust
irradiation by EUV photons from shock breakout under conditions where
the RSG wind is optically thin (a maximum H column of $1.5\times
10^{20}$ cm$^{-2}$). The hydrogen column density in the wind is
approximately $1.4\times 10^{38}/r_{bub}$ where $r_{bub}$ is the
radius of the inner edge of the RSG wind, which is the bubble
radius. For the H column of \citet{dwek08}, $r_{bub}$ is $\sim 0.3$
pc, which is entirely consistent with our previous discussion. At
higher H column densities, the required shock breakout luminosities
exceed $10^{12}L_{\sun}$. These are higher than those modeled by
\citet{blinnikov00} for the case of SN 1987A, so while they are not
ruled out, they must be considered unlikely.  Of course radiation from
the SN event is likely to completely ionize the surrounding CSM,
especially as there is no neutral material ahead of the forward shock
at 2.5 pc radius. Our concern here, however, is with the shock
breakout radiation, which represents a small fraction of the total
radiation in photons from the supernova event, and which is most
likely emitted {\em before} such photoionization takes place.

SN shock breakout has also been observed {\em in real time} by {\it
  Galaxy Evolution Explorer} \citep[GALEX;][]{ schawinski08,
  gezari08}. In each case, the rise in UV emission as the shock
emerges through the photosphere of the RSG progenitor was observed,
although the two sets of authors have slightly different
interpretations. By way of contrast, there are also claims that shock
breakout was observed in explosions of more stripped progenitors that
are more similar to Cas A, such as the Type Ib SN 2008D observed in
X-rays by {\it Swift}.  Early interpretations \citep{soderberg08,
  chevalier08} inferred a breakout radius of $\sim 6 \times 10^{11}$
cm, that was noted to be somewhat larger than the expected progenitor
radius of WR stars.  This would make the presence a dense stellar wind
close to the stellar surface likely, with no significant ``bubble''.
In their thorough study of the SN 2008D observations, however,
\citet{modjaz09} show that the radius inferred from observations is in
fact consistent with that of the expected WN progenitor radius within
errors.  There is then no need for a dense wind close to the
progenitor, but the distance that would have been involved is in any
case substantially smaller than the putative bubble radius for Cas A.
Claims of shock breakout in SN 2006aj are more controversial
\citep[e.g.,][]{soderberg06,chevalier08}.

It is also worthwhile to consider the similarities and differences
between Cas A and the long-soft GRBs that are associated with
core-collapse supernovae.  While Cas A was not likely associated with
a ``classical'' gamma-ray burst \citep{laming06}, it might have some
relation to lower energy GRBs or X-ray flashes (XRFs).  Its progenitor
clearly underwent the same kind of substantial mass loss that allows
the relativistic jet in GRBs to penetrate the stellar layers at
explosion and generate the burst. Further similarities include the
presence of Cas A's ejecta jets, and an inferred mass at explosion and
explosion energy that are in line with those inferred for the less
energetic examples of long GRBs. In particular, the explosion energy
and progenitor mass at explosion determined by \citet{laming03} for
Cas A are nearly identical to the values of 2$\times 10^{51}$ ergs and
2 M$_\odot$, respectively, inferred for GRB/XRF 060218/SN 2006aj
\citep{mazzali06}, though this Type Ic event was even more stripped by
stellar wind mass loss than was Cas A. Similar mass and explosion
energies have also been inferred for other examples of unusual Ib/Ic
explosions that are not associated with GRBs, such as SN 2008D
\citep[ejecta mass 3-5 M$_{\sun}$, explosion energy $2-4\times
  10^{51}$ erg,][]{soderberg08} and SN 2005bf \citep[ejecta mass 8.3
  M$_\odot$, explosion energy $2\times 10^{51}$ erg,][]{folatelli06}.

\section{Conclusion}

In this work, we have explored a simplified one-dimensional
circumstellar environment that nevertheless provides a workable model
for Cas A. We have presented several lines of evidence to suggest that
Cas A evolved into a small circumstellar bubble of approximately
0.2-0.3 pc radius located inside the circumstellar wind.  Aside from
allowing higher temperatures for the ejecta knots that are more
consistent with the results of spectral fits to X-ray data, the
presence of a bubble provides better agreement with the dynamics and
radii of the shocks, as noted previously, and reduces the mass of
radiatively-cooled ejecta to be in better agreement with the optical
observations.  Further support for the the likely presence of
circumstellar bubbles in Cas A include its likely binary progenitor,
and estimates for the supernova shock breakout luminosity and its
processing in the surrounding environment.

One puzzle that remains here involves the presence of nonthermal
emission superposed with the ejecta knot spectra, and the extent to
which these might affect inferred temperatures for the ejecta knots.
The temperatures seen in the spectra are systematically higher exactly
where the nonthermal emission is most prominent, but a more
sophisticated and larger-scale treatment of the nonthermal emission
may be required to assess this properly, as it is difficult to
constrain the low level of nonthermal emission in these ejecta spectra.

Finally, the ejecta spectra studied here clearly show the presence of
a component that is most simply described by plasma highly enriched in
Fe.  This component occurs in knots at locations throughout Cas A that
we have studied here, and appears to tend toward higher temperatures
and ionization ages.  The presence of highly pure Fe implies explosive
Si burning, possibly by alpha-rich freezeout.  Such highly pure Fe was
identified by Chandra in the southeast \citep{hwang03}, but the
present study indicates that such pure Fe may be present on a more
extensive scale.  The distinct thermodynamic parameters of the
additional Fe component suggests that it is a separate emission
component that is projected onto the line of sight together with
ejecta of a more normal composition.  A more comprehensive study of
the Fe ejecta in Cas A, however, is required to draw firm conclusions.

\acknowledgments UH and JML acknowledge support through NASA LTSA
grant NNG06GB89G.  JML was also supported by basic research funds of
the Office of Naval Research.  We thank Rober Chevalier and the
anonymous referee for helpful comments on the paper.

\begin{figure}
\centerline{\hspace{0.4in}\includegraphics[scale=0.38]{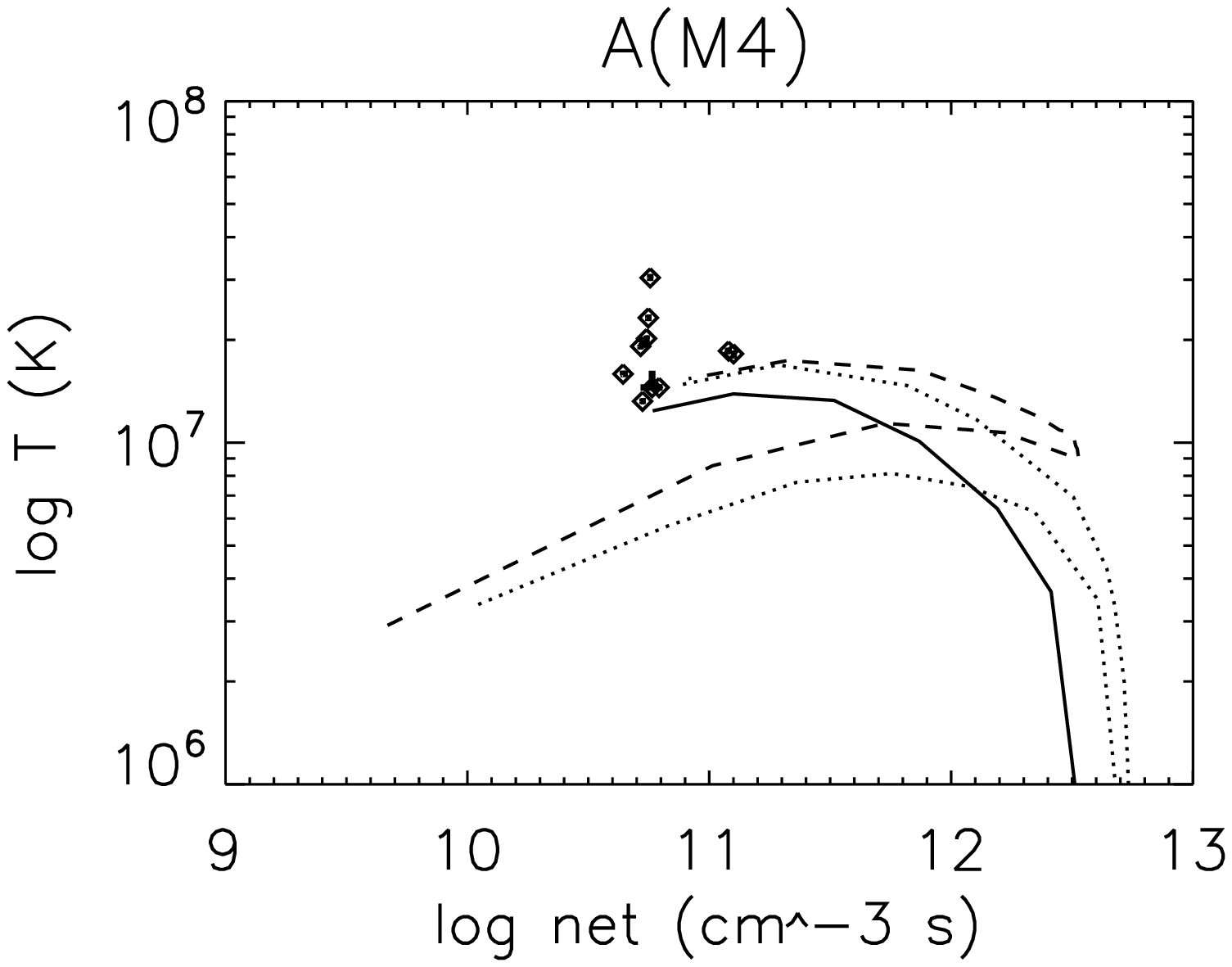}\includegraphics[scale=0.38]{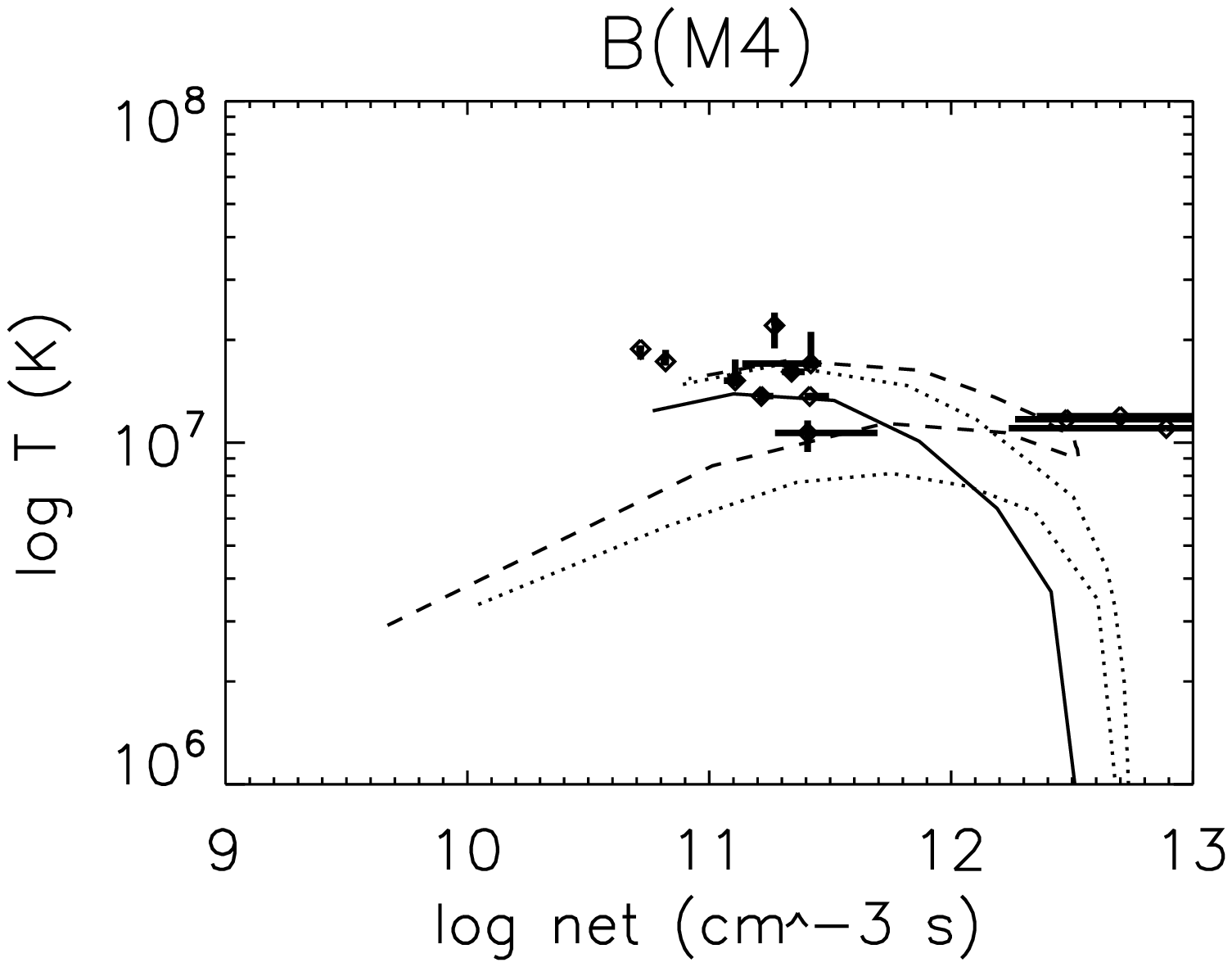}\includegraphics[scale=0.38]{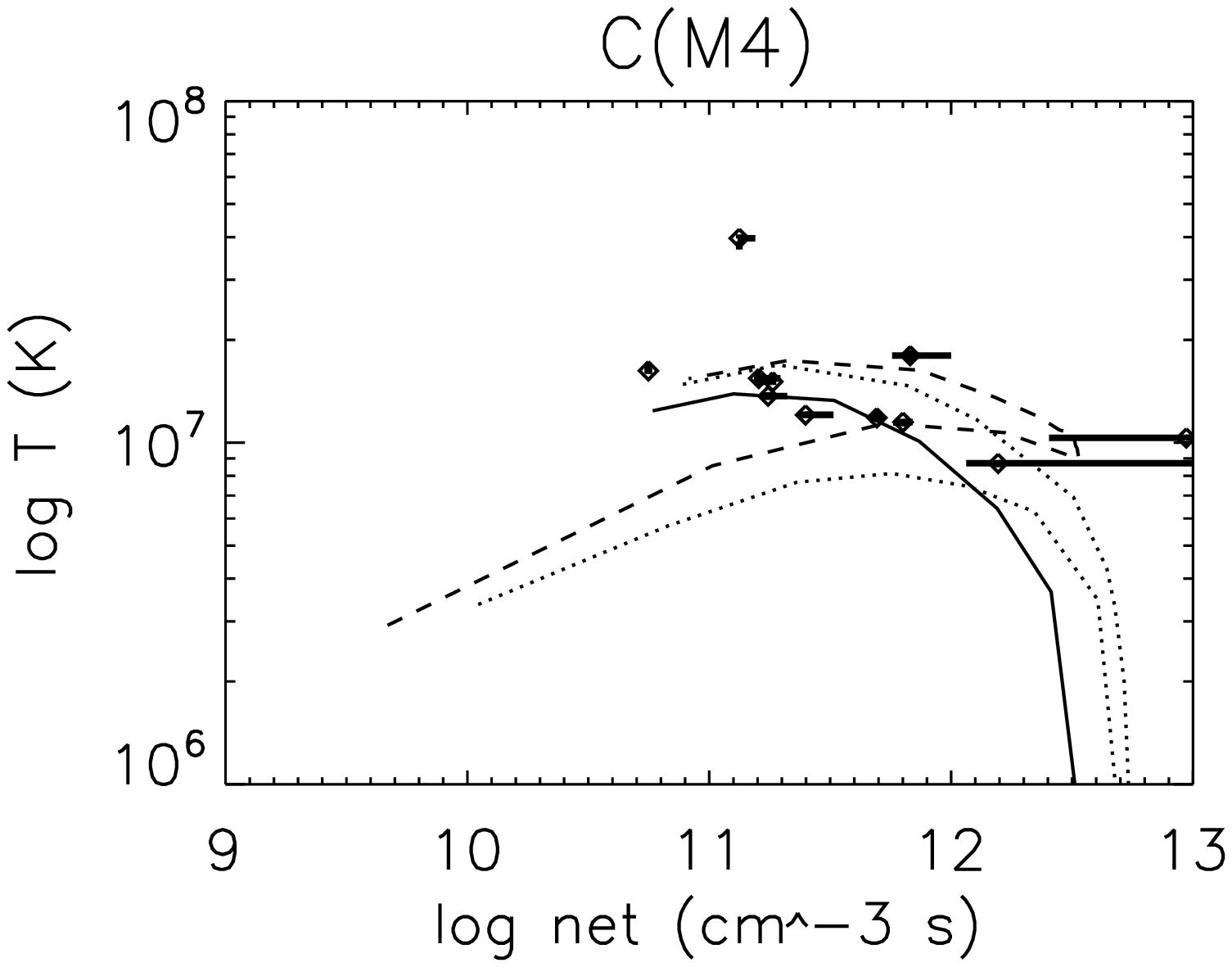}}
\centerline{\hspace{0.4in}\includegraphics[scale=0.38]{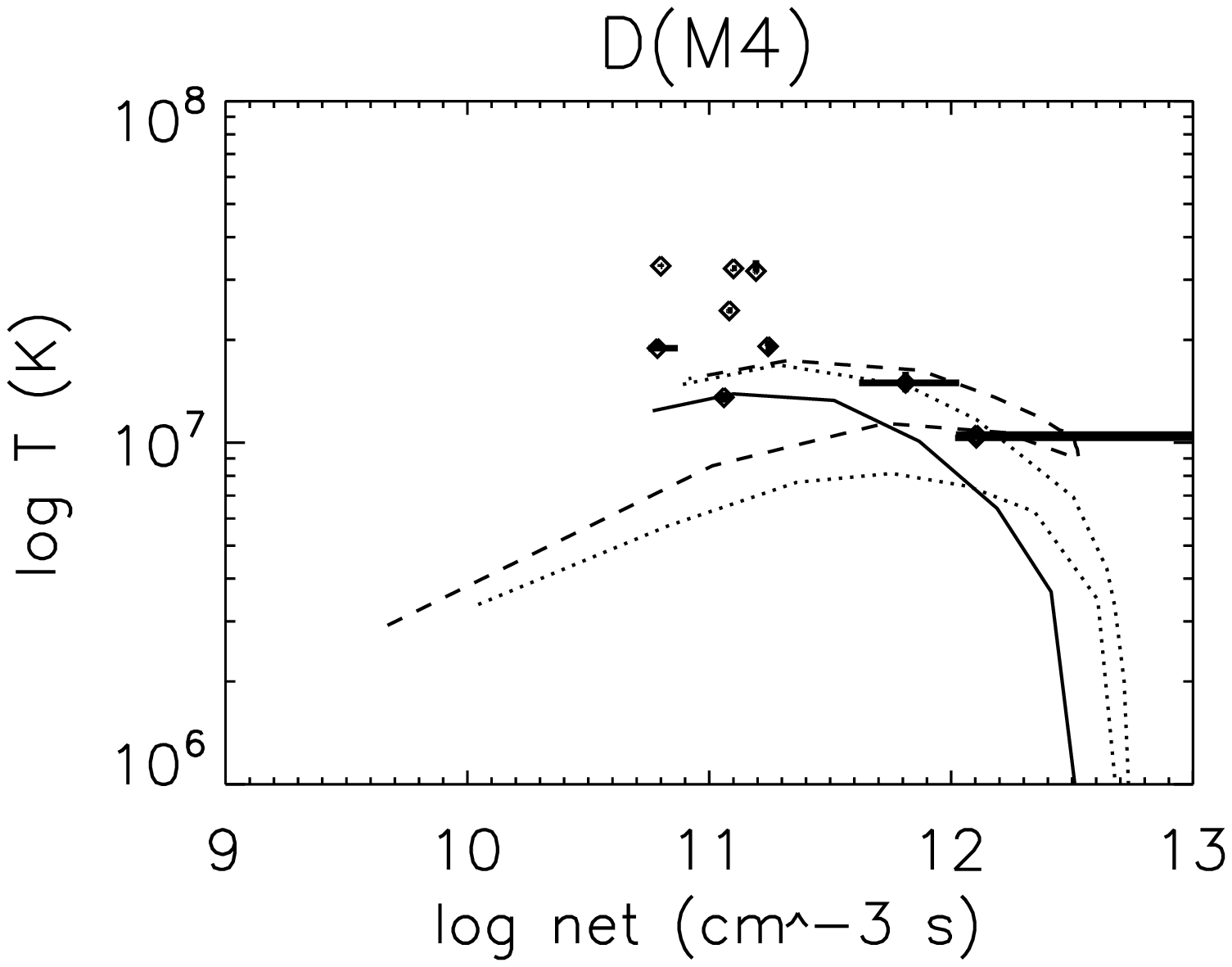}\includegraphics[scale=0.38]{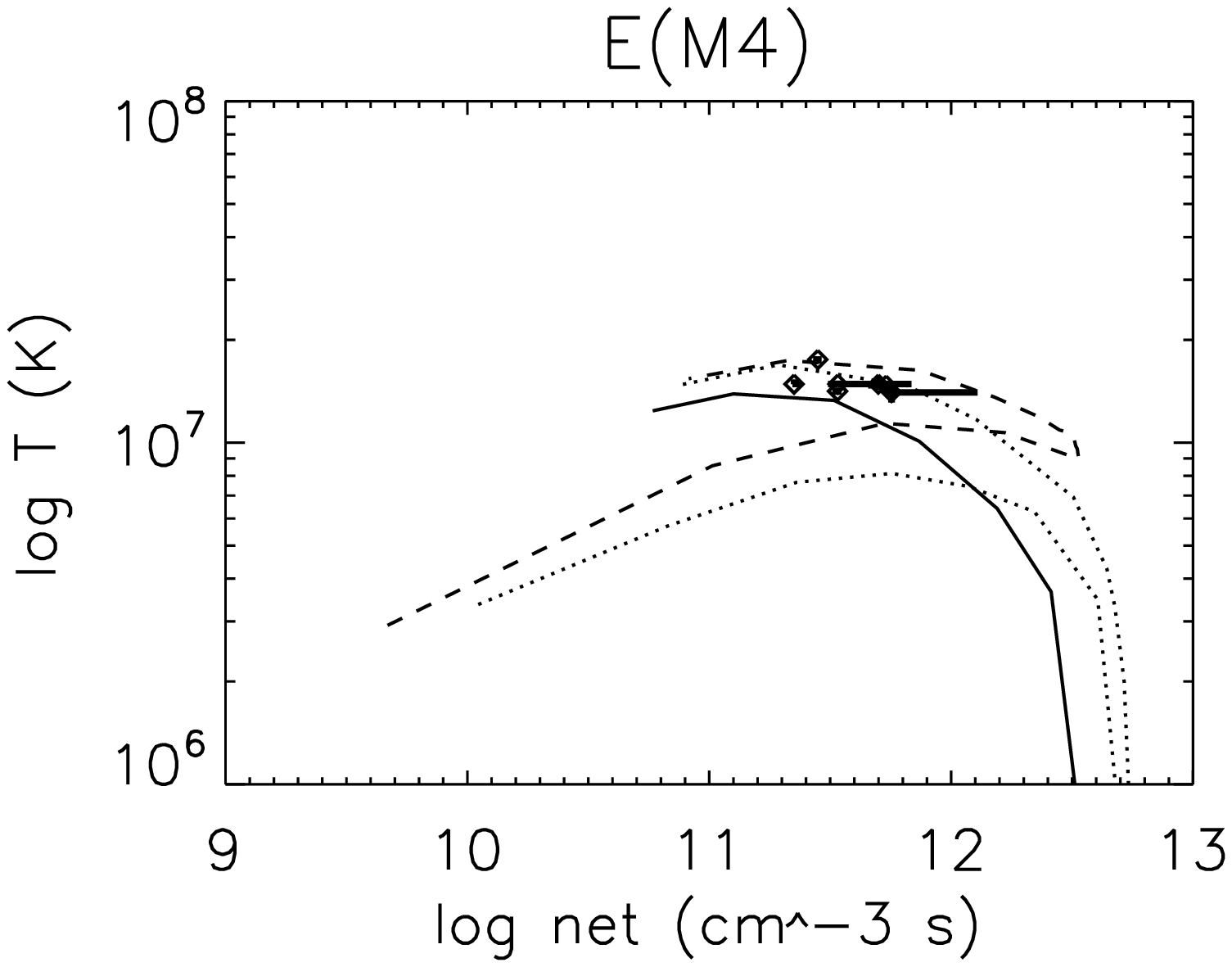}\includegraphics[scale=0.38]{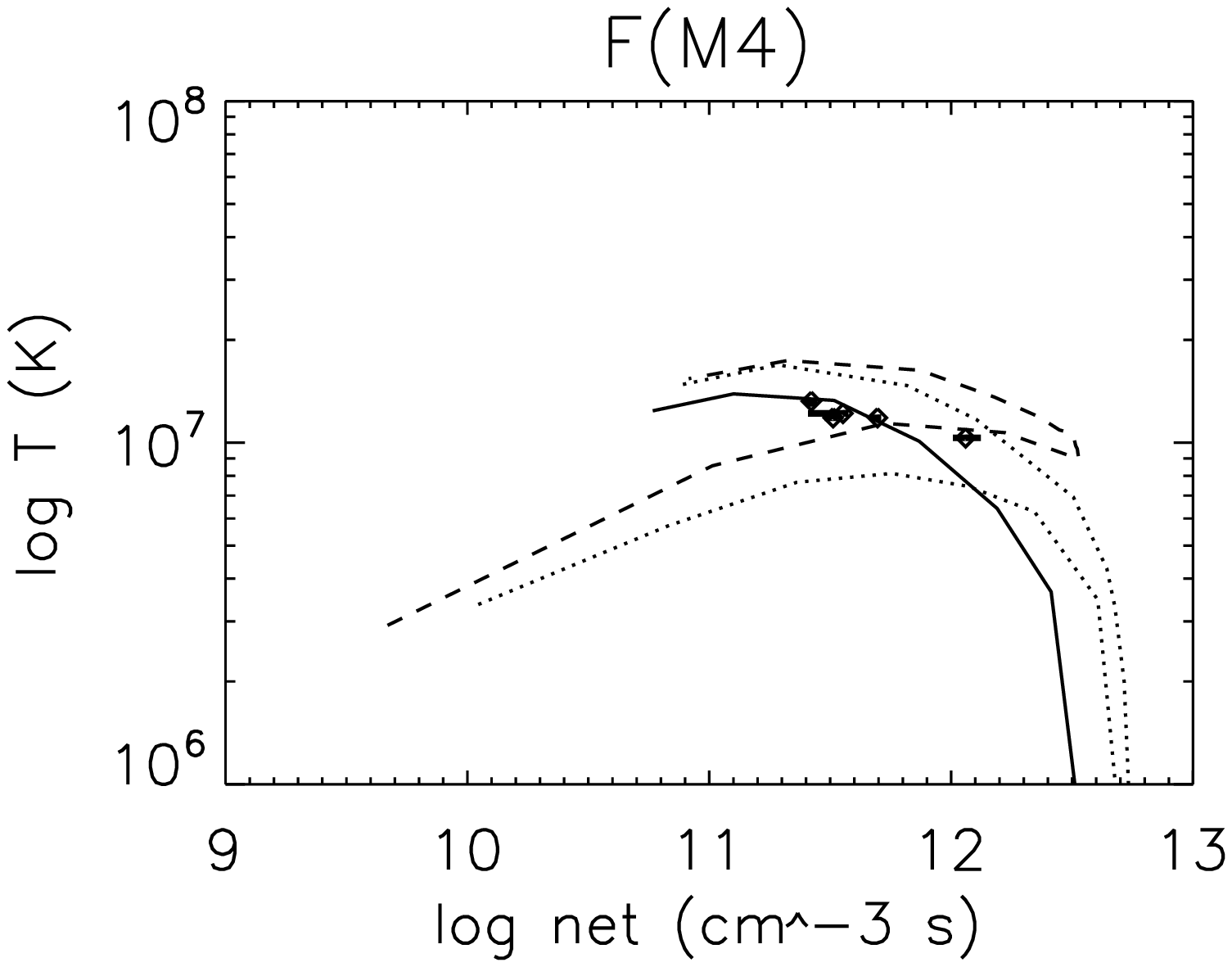}}
\centerline{\hspace{0.4in}\includegraphics[scale=0.38]{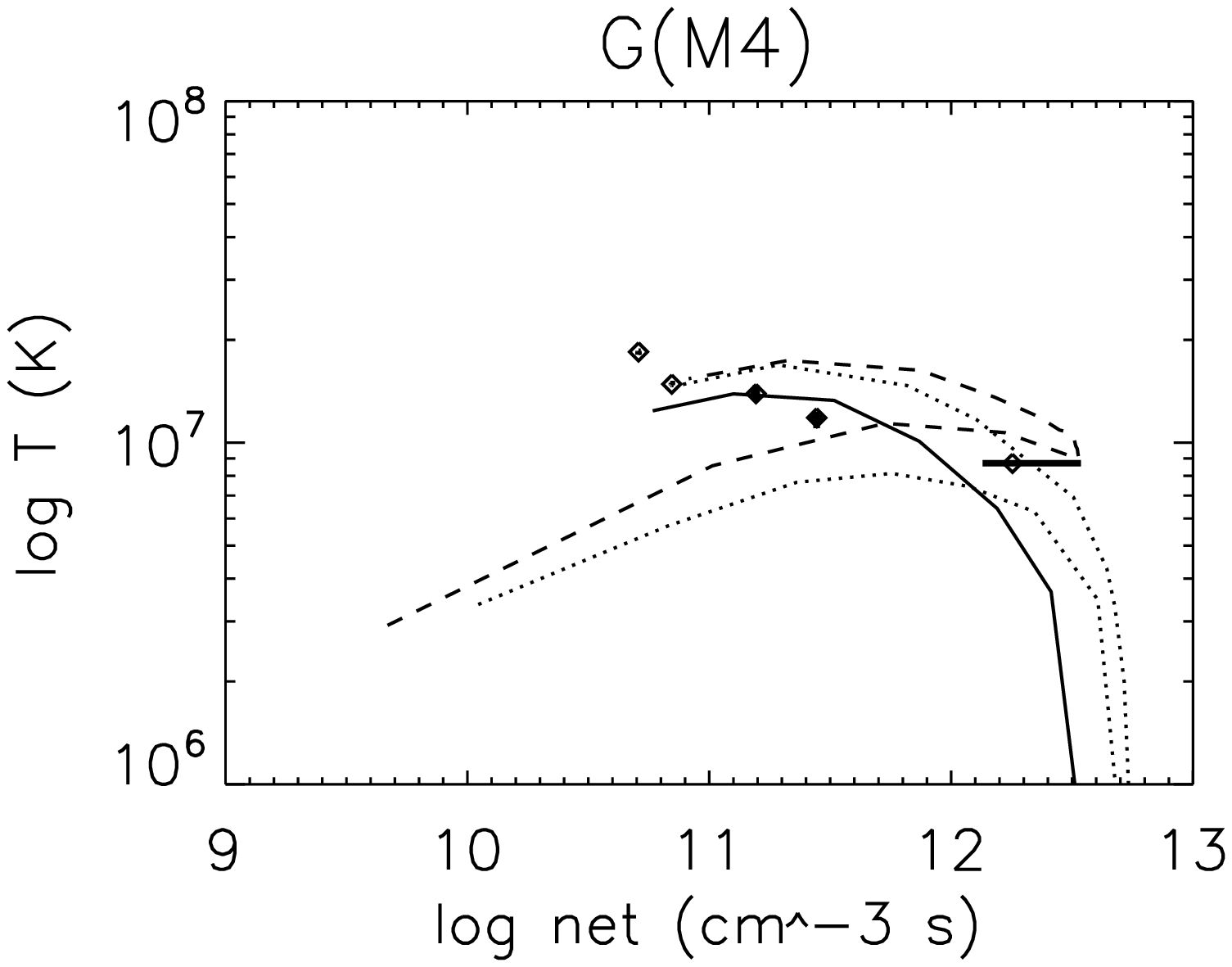}\includegraphics[scale=0.38]{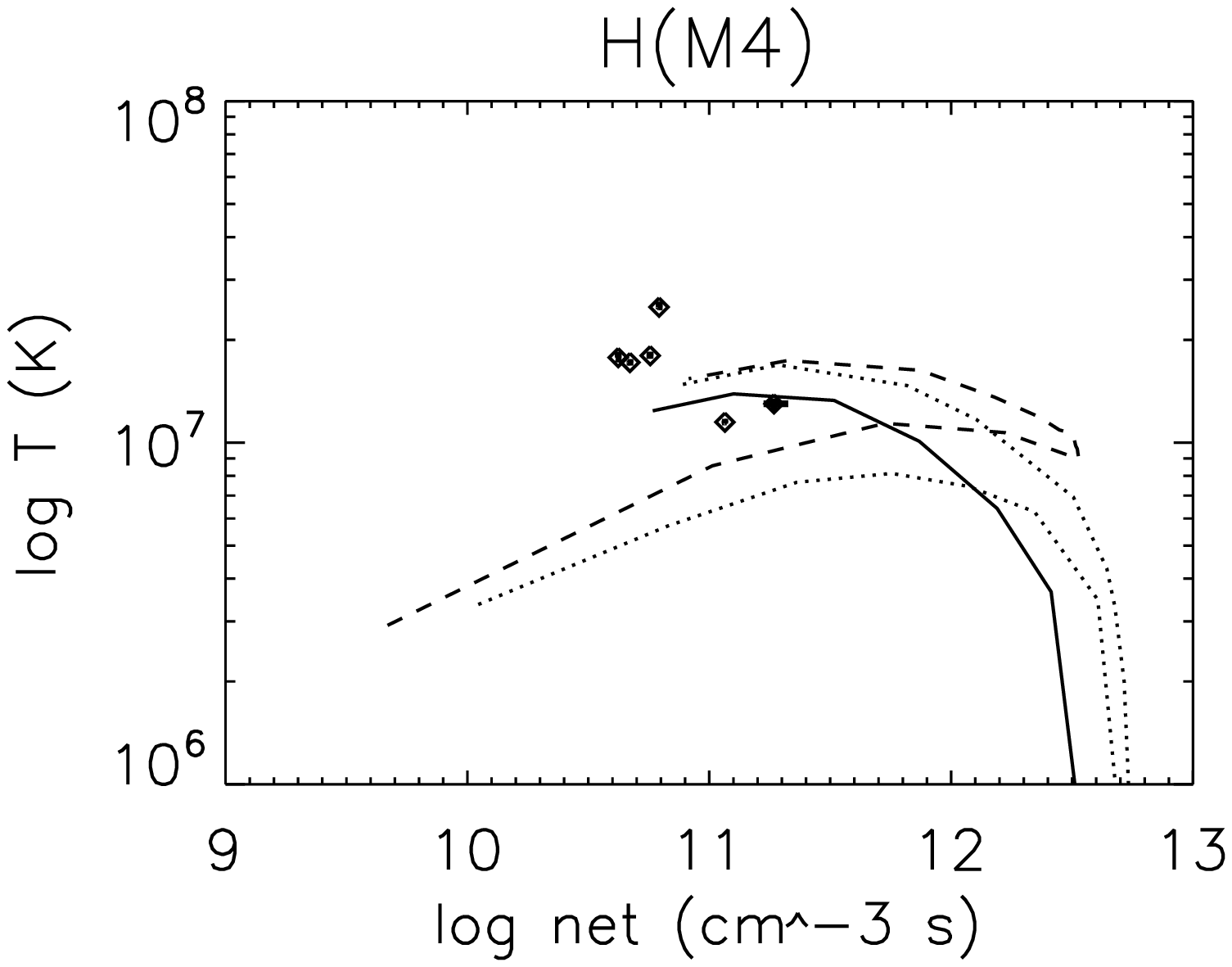}\includegraphics[scale=0.38]{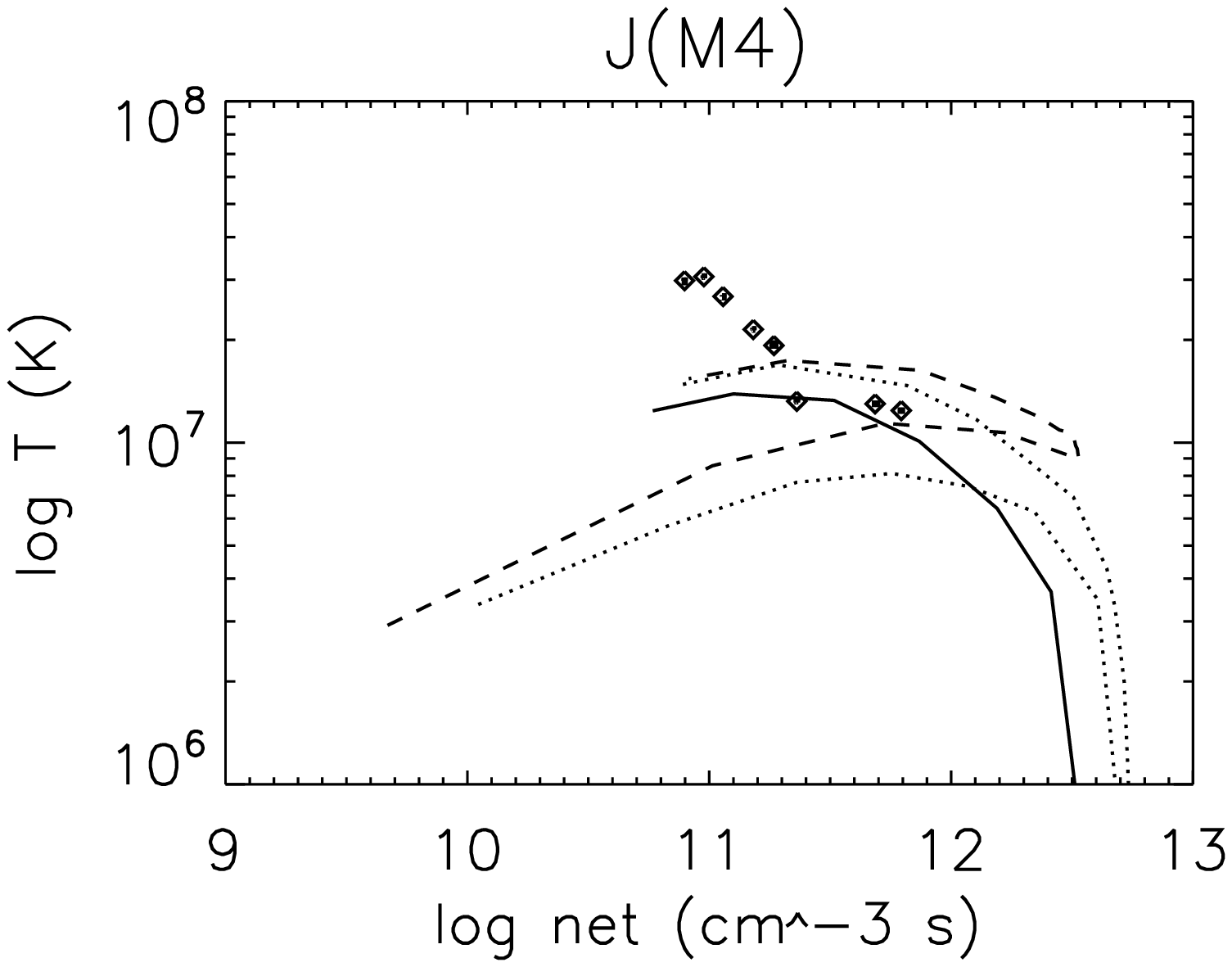}}

\hspace{0.8in}\caption{Plots of measured temperature and ionization age for each of
  the series of knots pictured in Figure \ref{knots} for the M4 set of
  abundances in Table 2.  The curves show models for the evolution of
  temperature and ionization age in a circumstellar wind with a
  central bubble of size 0, 0.2, and 0.3 pc respectively (bottom to
  top: solid, dotted, and dashed).  For series N, models for the K10
  and C10 sets of abundances are also shown, to illustrate the effect
  of the range of abundances considered in the calculated models.}
\label{tnet}
\end{figure}

\begin{figure}
\setcounter{figure}{2}
\centerline{\hspace{0.4in}\includegraphics[scale=0.38]{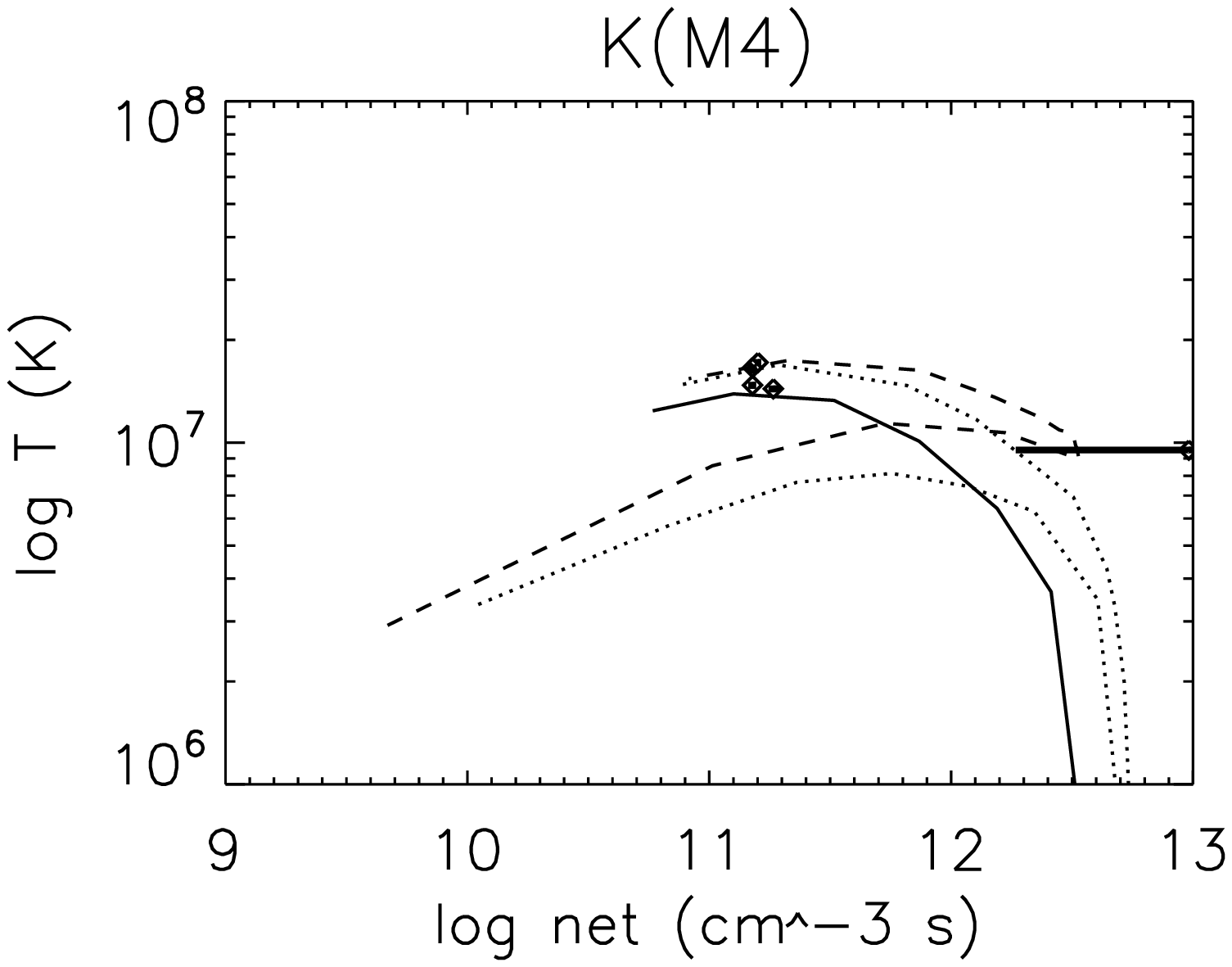}\includegraphics[scale=0.38]{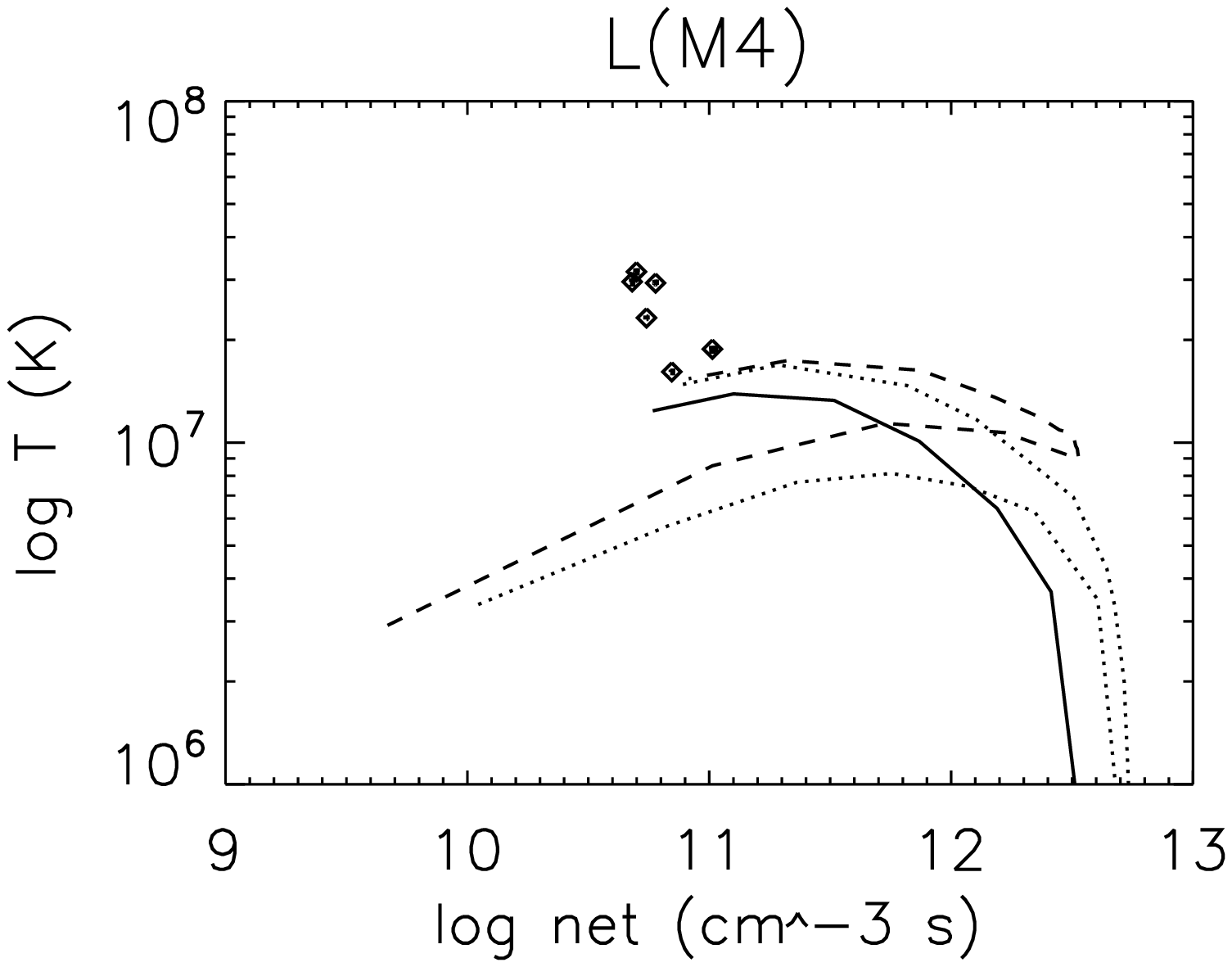}\includegraphics[scale=0.38]{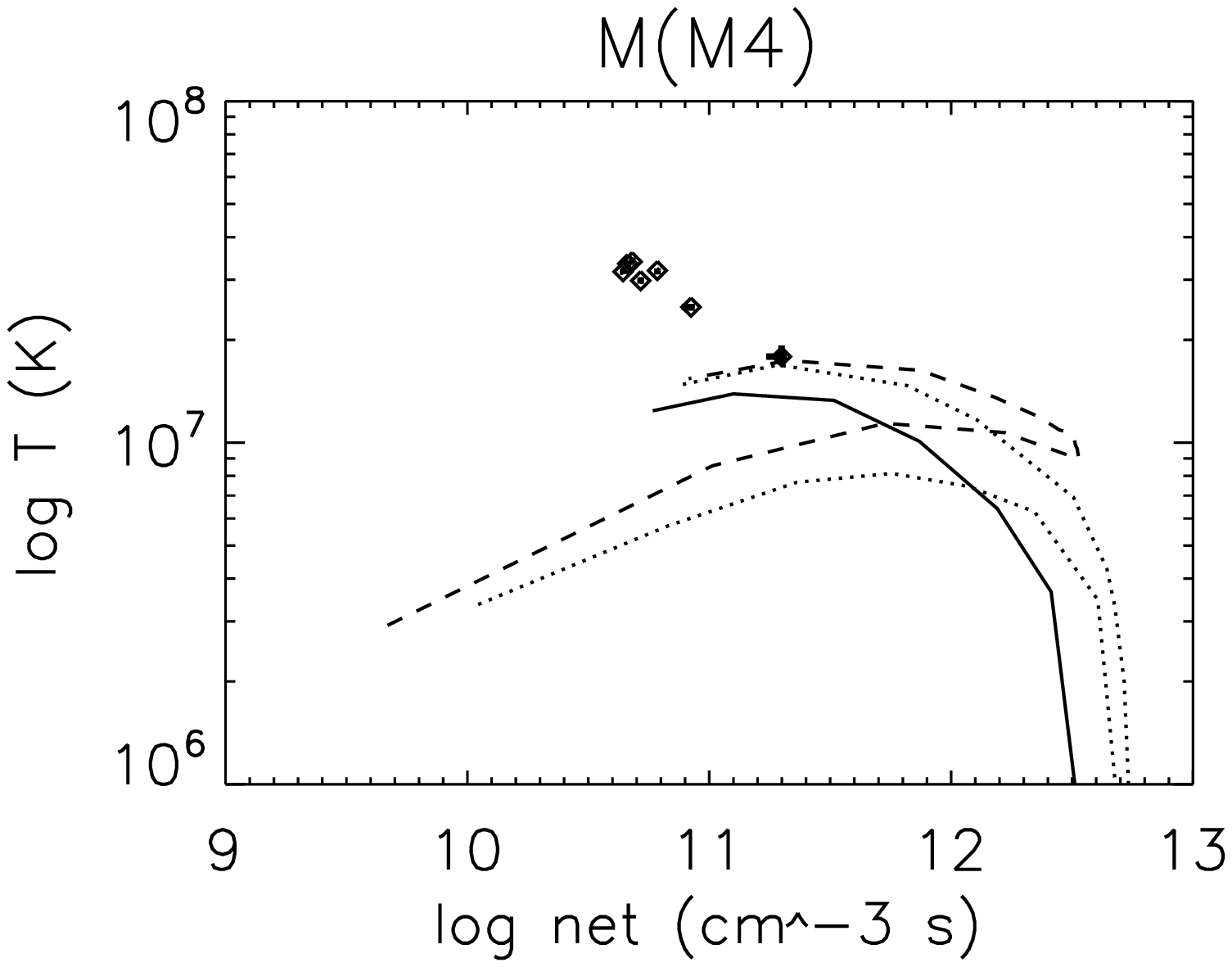}}
\centerline{\hspace{0.4in}\includegraphics[scale=0.38]{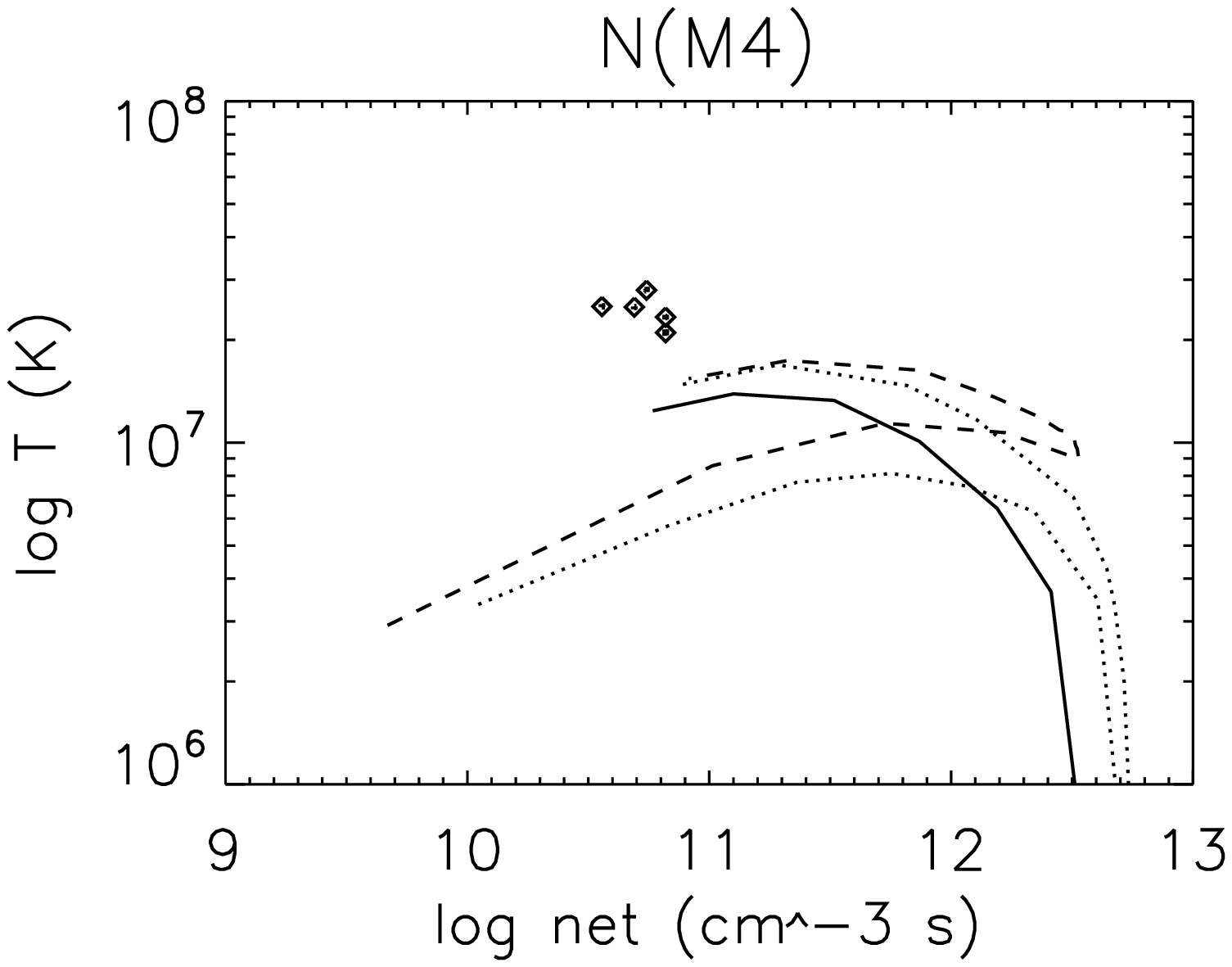}\includegraphics[scale=0.38]{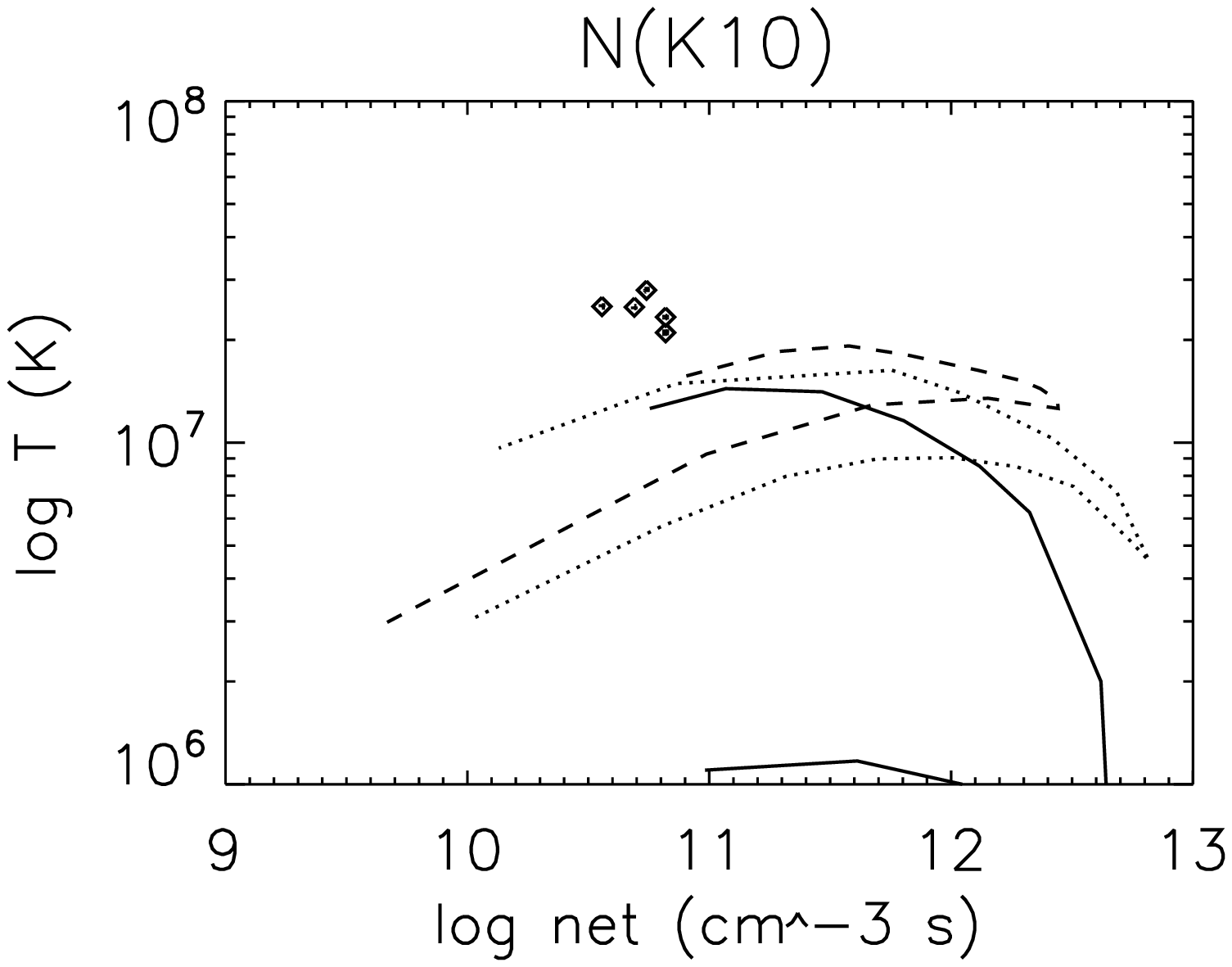}\includegraphics[scale=0.38]{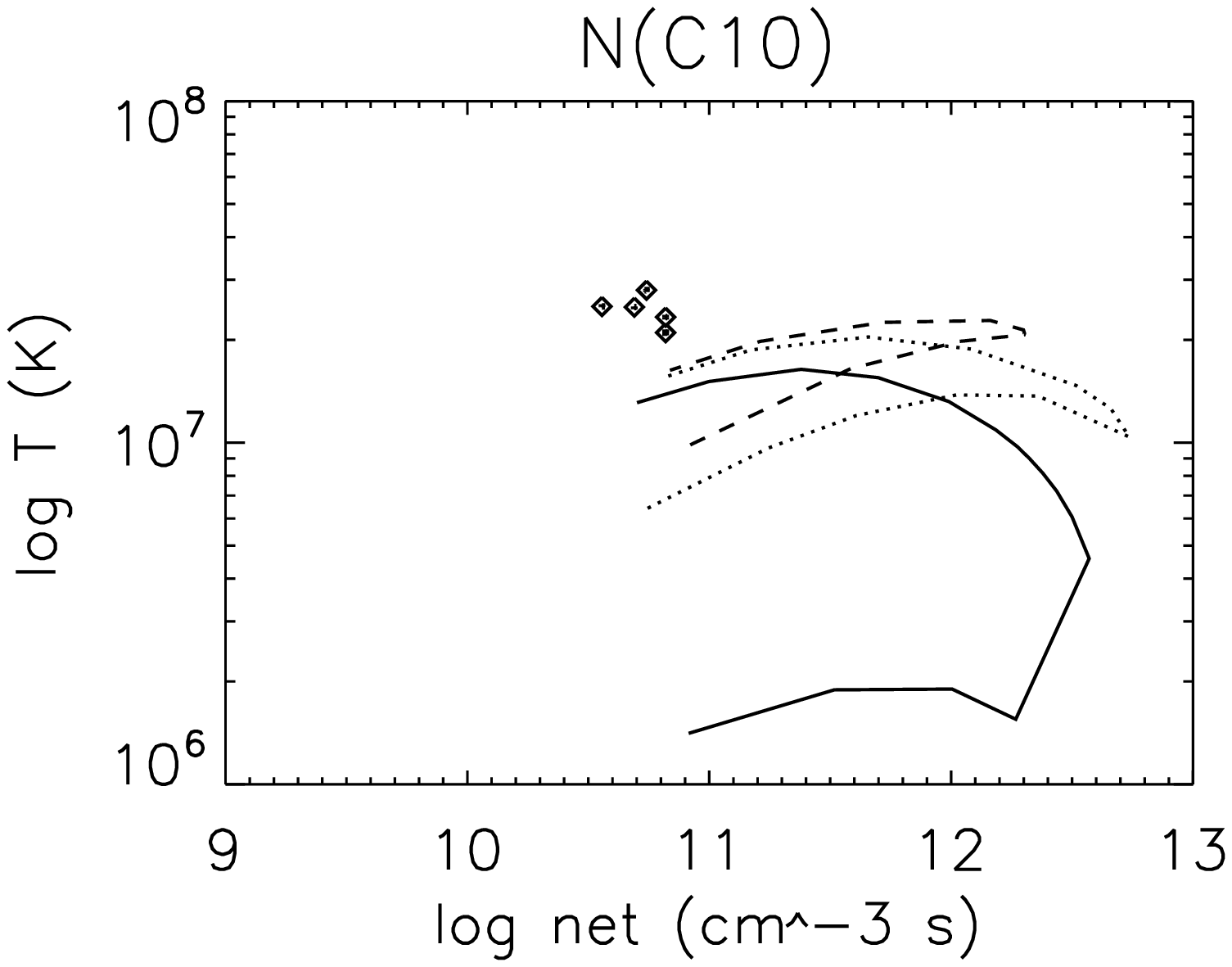}}
\hspace{0.8in}\caption{continued}
\end{figure}

\setcounter{table}{1}
\begin{deluxetable}{cccc}
\tabletypesize{\footnotesize}
\tablecaption{SNR Models}
\tablewidth{0pt}
\tablehead{
\colhead{Bubble size (pc)}&\colhead{blast wave radius (pc)}&
\colhead{blast wave speed (km s$^{-1}$)}&\colhead{SNR age (yrs)}}
\startdata
0.0 & 2.37& 5072 & 325\\
0.0 & 2.5 & 4930 & 350\\
0.1 & 2.5 & 4791 & 339\\
0.2 & 2.5 & 5007 & 331\\
0.3 & 2.5 & 5044 & 324\\
\enddata
\end{deluxetable}

\begin{deluxetable}{ccccccccc}
\tabletypesize{\footnotesize}
\tablecaption{Abundance sets by mass}
\tablewidth{0pt}
\tablehead{
\colhead{knot}&\colhead{O}&
\colhead{Ne}&\colhead{Mg}&\colhead{Si}&\colhead{S}&\colhead{Ar}&\colhead{Ca}& \colhead{Fe}}
\startdata
M4 & 0.941& 0.006   &       & 0.017 & 0.009 & 0.003 & 0.002 & 0.022\\
M6 & 0.9092 & 0.0321  & 0.004 & 0.0306& & & & 0.0281\\
K4 & 0.783  & 0.187   &       &       & & & &0.030\\
K10& 0.9260 & 0.0485  &       &  0.0143 & & & &0.0112\\
G3 & 0.9185 &         &       & 0.0408& & & & 0.0407\\
C10& 0.711  & 0.1413  &       & 0.0875& & & & 0.0602\\
\enddata
\end{deluxetable}

\eject
\setcounter{table}{0}
\hoffset=-0.85in
\begin{deluxetable}{cccccccccc}
\tabletypesize{\tiny}
\tablecaption{Two NEI Component Spectral Fits to Ejecta Knots}
\tablewidth{8in}
\tablehead{
\colhead{Knot}&\colhead{$\chi^2$}&\colhead{$N_H$}&\colhead{$kT$}&\colhead{$n_et$}&\colhead{Si}&\colhead{Fe} &\colhead{$kT_{Fe}$}&\colhead{$n_et_{Fe}$} \\ 
\colhead{}&\colhead{}&\colhead{(cm$^{-2}$)}&\colhead{(keV)}&\colhead{(cm$^{-3}$s)}&\colhead{(rel $\odot$)}&\colhead{(rel $\odot$)} &\colhead{(keV)}&\colhead{(cm$^{-3}$s)} \\ }
\startdata
A1 & 316.0, 1.43 & 1.45 (1.45-1.47) & 1.25 (1.23-1.27) & 0.62 (0.61-0.64) & 0.73 (0.72-0.75) & 0.05 (0.05-0.06) &68.80 (69.62-67.98) & 0.00 (0.00-0.02) \\
A2 & 357.5, 1.54 & 1.53 (1.52-1.54) & 1.14 (1.12-1.16) & 0.53 (0.52-0.54) & 0.90 (0.88-0.91) & 0.05 (0.05-0.06) & 1.64 (1.51-1.79) &38.39 (7.51-500) \\
A3 & 265.9, 1.01 & 1.36 (1.35-1.37) & 2.62 (2.57-2.69) & 0.57 (0.56-0.58) & 0.47 (0.46-0.48) & 0.06 (0.06-0.07) & 2.67 (2.49-3.99) & 2.51 (2.28-3.77) \\
A4 & 162.4, 0.76 & 1.38 (1.34-1.44) & 1.25 (1.23-1.40) & 0.58 (0.52-0.61) & 0.53 (0.48-0.56) & 0.07 (0.07-0.08) & 5.13 (3.75-8.41) & 1.03 (0.97-1.37) \\
A5 & 411.6, 1.60 & 1.64 (1.62-1.65) & 1.37 (1.36-1.40) & 0.44 (0.43-0.46) & 0.70 (0.67-0.72) & 0.01 (0.01-0.02) & 0.92 (0.89-0.96) & 310.04 (4.06-500) \\
A6 & 370.5, 1.17 & 1.52 (1.51-1.55) & 1.74 (1.67-1.77) & 0.55 (0.54-0.56) & 0.15 (0.15-0.16) & 0.06 (0.06-0.07) & 2.28 (2.18-2.54) & 2.87 (2.66-3.33) \\
A7 & 199.4, 0.68 & 1.24 (1.23-1.25) & 1.60 (1.57-1.62) & 1.20 (1.16-1.22) & 0.11 (0.11-0.11) & 0.02 (0.02-0.02) & 1.76 (1.71-1.81) & 0.29 (0.27-0.30) \\
A8 & 296.2, 1.22 & 1.38 (1.38-1.39) & 1.65 (1.62-1.67) & 0.52 (0.51-0.53) & 0.59 (0.58-0.60) & 0.13 (0.13-0.13) & 1.81 (1.73-1.91) &11.28 (7.54-500) \\
A9 & 333.8, 1.28 & 1.41 (1.40-1.42) & 2.00 (1.96-2.04) & 0.56 (0.55-0.57) & 0.60 (0.59-0.61) & 0.12 (0.11-0.12) & 1.89 (1.79-2.02) &28.78 (7.19-500) \\
A10& 241.7, 0.97 & 1.16 (1.15-1.18) & 1.57 (1.52-1.59) & 1.26 (1.23-1.29) & 0.45 (0.45-0.47) & 0.04 (0.04-0.04) & 4.76 (4.70-4.76) & 0.01 (0.00-0.01) \\
B1 & 149.5, 0.77 & 1.36 (1.35-1.38) & 1.62 (1.51-1.66) & 0.52 (0.50-0.53) & 0.82 (0.80-0.84) & 0.21 (0.20-0.22) & 1.82 (1.72-1.88) & 4.57 (4.11-5.15) \\
B2 & 175.9, 0.87 & 1.36 (1.34-1.40) & 1.49 (1.45-1.61) & 0.66 (0.63-0.67) & 0.66 (0.65-0.73) & 0.23 (0.21-0.24) & 3.35 (3.17-3.96) & 1.54 (1.47-2.04) \\
B3 & 182.3, 0.88 & 1.10 (1.04-1.13) & 0.92 (0.81-1.00) & 2.55 (1.87-4.96) & 0.70 (0.62-1.06) & 0.08 (0.07-0.13) & 2.11 (1.46-2.52) &41.63 (3.72-500) \\
B4 & 152.8, 0.74 & 1.47 (1.45-1.51) & 1.31 (1.26-1.51) & 1.28 (1.15-1.36) & 0.69 (0.63-0.78) & 0.20 (0.18-0.22) & 0.15 (0.14-0.17) & 0.12 (0.10-5.21) \\
B5 & 221.9, 0.96 & 1.37 (1.36-1.41) & 1.18 (1.13-1.24) & 1.64 (1.58-1.84) & 0.56 (0.55-0.59) & 0.18 (0.14-0.20) & 1.39 (1.37-1.42) & 0.18 (0.15-0.21) \\
B6 & 160.3, 0.75 & 1.16 (1.14-1.17) & 1.18 (1.16-1.19) & 2.60 (2.48-3.12) & 0.41 (0.36-0.42) & 0.11 (0.10-0.11) & 2.34 (2.21-2.54) & 3.88 (3.46-4.62) \\
B7 & 153.7, 0.69 & 1.56 (1.49-1.60) & 1.39 (1.32-1.48) & 2.19 (2.00-2.48) & 0.41 (0.40-0.45) & 0.31 (0.28-0.36) & 1.36 (0.89-1.71) & 0.15 (0.10-0.21) \\
B8 & 168.4, 0.72 & 1.06 (0.99-1.10) & 1.90 (1.63-2.07) & 1.86 (1.81-2.01) & 0.40 (0.39-0.42) & 0.16 (0.15-0.19) &79.90 (0.00-53.58) & 1.02 (0.93-1.09) \\
B9 & 195.8, 0.94 & 1.09 (1.07-1.13) & 0.95 (0.94-0.97) &77.49 (17.30-500) & 0.50 (0.47-0.52) & 0.07 (0.07-0.08) & 2.05 (1.96-2.18) & 5.57 (5.06-6.31) \\
B10& 140.3, 0.68 & 1.58 (1.53-1.62) & 1.47 (1.42-1.82) & 2.63 (1.37-2.90) & 0.60 (0.51-0.63) & 0.62 (0.52-0.67) & 0.64 (0.59-0.69) &20 (1.75-500) \\
B11& 200.0, 0.85 & 1.40 (1.38-1.43) & 1.03 (1.03-1.04) &50 (22.60-500) & 0.81 (0.66-0.84) & 0.42 (0.38-0.44) & 2.50 (2.40-2.71) & 2.19 (2.03-2.37) \\
B12& 204.8, 1.00 & 1.13 (1.12-1.18) & 1.01 (0.99-1.01) &30 (18.40-500) & 0.37 (0.34-0.41) & 0.14 (0.14-0.15) & 2.51 (2.43-2.97) & 2.30 (2.20-2.56) \\
C1 & 181.7, 0.82 & 1.51 (1.50-1.54) & 1.40 (1.37-1.48) & 0.56 (0.54-0.58) & 0.62 (0.59-0.66) & 0.14 (0.13-0.15) & 1.52 (1.46-1.59) &34.09 (10.46-500) \\
C2 & 184.8, 0.88 & 1.14 (1.13-1.16) & 1.04 (1.02-1.05) & 2.50 (2.35-3.26) & 0.46 (0.45-0.51) & 0.05 (0.04-0.05) & 2.34 (2.26-2.45) & 4.46 (4.04-12.96) \\
C3 & 196.3, 0.86 & 1.17 (1.16-1.18) & 1.02 (1.00-1.07) & 4.93 (4.61-5.32) & 0.43 (0.42-0.44) & 0.06 (0.06-0.07) & 2.15 (1.18-2.24) & 5.02 (4.59-5.64) \\
C4 & 208.2, 0.91 & 0.98 (0.96-1.02) & 1.18 (1.15-1.21) & 1.75 (1.64-2.10) & 0.38 (0.37-0.41) & 0.03 (0.03-0.03) & 2.36 (2.28-2.47) & 5.89 (5.17-7.65) \\
C5 & 282.4, 1.10 & 1.81 (1.79-1.83) & 1.30 (1.28-1.38) & 1.83 (1.60-1.89) & 0.35 (0.34-0.36) & 0.17 (0.16-0.18) & 0.68 (0.62-0.80) & 0.43 (0.37-0.53) \\
C6 & 274.9, 1.16 & 1.16 (1.15-1.18) & 0.99 (0.98-1.00) & 6.32 (5.84-6.93) & 0.37 (0.36-0.41) & 0.03 (0.03-0.04) & 2.61 (2.54-2.71) & 3.60 (3.43-3.83) \\
C7 & 223.4, 0.98 & 1.99 (1.89-2.01) & 1.33 (1.26-1.40) & 1.60 (1.53-1.97) & 0.55 (0.50-0.62) & 0.24 (0.15-0.26) & 0.54 (0.50-0.64) & 1.56 (0.61-500) \\
C8 & 268.7, 1.16 & 1.83 (1.82-1.85) & 0.89 (0.88-0.90) &93.83 (25.40-500) & 0.93 (0.90-1.12) & 0.58 (0.46-0.60) & 2.30 (2.23-2.38) & 2.67 (2.53-2.82) \\
C9 & 240.9, 0.97 & 0.77 (0.75-0.78) & 3.42 (3.17-3.51) & 1.33 (1.31-1.55) & 0.46 (0.44-0.54) & 0.20 (0.18-0.21) &54.63 (49.60-58.99) & 0.95 (0.89-1.00) \\
C10& 302.3, 1.15 & 0.88 (0.86-0.89) & 1.55 (1.48-1.61) & 6.77 (5.70-10) & 0.54 (0.52-0.55) & 0.18 (0.18-0.19) &52.28 (40.68-57.52) & 0.95 (0.90-1.02) \\
C11& 174.9, 0.85 & 0.98 (0.93-1.02) & 0.75 (0.73-0.76) &15.64 (11.54-500) & 0.54 (0.52-0.64) & 0.07 (0.06-0.08) &10.15 (10.08-10.31) & 0.86 (0.80-0.87) \\
D1 & 380.8, 1.40 & 1.21 (1.18-1.22) & 1.17 (1.13-1.21) & 1.15 (1.06-1.24) & 0.44 (0.43-0.45) & 0.03 (0.03-0.03) & 2.14 (2.08-2.22) &10.36 (7.69-500) \\
D2 & 342.4, 1.28 & 1.46 (1.44-1.47) & 1.63 (1.60-1.72) & 0.61 (0.56-0.74) & 0.54 (0.52-0.56) & 0.08 (0.07-0.08) & 1.86 (1.83-1.93) &19.99 (8.93-500) \\
D3 & 850.3, 2.39 & 1.56 (1.55-1.57) & 2.84 (2.82-2.87) & 0.63 (0.63-0.64) & 0.28 (0.27-0.28) & 0.04 (0.04-0.04) & 2.08 (2.03-2.24) & 7.90 (6.14-500) \\
D4 & 335.6, 1.15 & 1.06 (1.05-1.07) & 1.65 (1.57-1.72) & 1.75 (1.71-1.86) & 0.33 (0.32-0.33) & 0.03 (0.03-0.03) & 2.62 (2.52-3.35) & 4.88 (4.38-5.71) \\
D5 & 658.8, 1.65 & 1.23 (1.22-1.24) & 2.79 (2.74-2.79) & 1.26 (1.24-1.26) & 0.28 (0.28-0.28) & 0.05 (0.05-0.05) &34.02 (26.38-40.67) & 0.97 (0.92-1.03) \\
D6 & 251.9, 0.87 & 1.09 (1.08-1.11) & 2.74 (2.71-2.95) & 1.56 (1.53-1.59) & 0.38 (0.37-0.39) & 0.06 (0.06-0.06) & 7.19 (4.54-9.92) &29.95 (2.56-500) \\
D7 & 249.9, 1.00 & 1.11 (1.09-1.14) & 1.29 (1.21-1.39) & 6.48 (4.17-10.78) & 0.49 (0.44-0.55) & 0.03 (0.02-0.04) & 3.46 (3.32-3.61) & 2.37 (2.26-4.64) \\
D8 & 204.3, 0.90 & 1.16 (1.11-1.19) & 0.89 (0.88-0.90) &12.70 (10.38-500) & 0.69 (0.64-0.74) & 0.04 (0.04-0.06) & 2.12 (1.79-2.19) & 5.19 (4.90-8.52) \\
D9 & 168.2, 0.82 & 1.08 (1.06-1.09) & 0.91 (0.90-0.92) &12.72 (10.47-500) & 0.60 (0.58-0.62) & 0.05 (0.04-0.05) & 2.37 (2.32-2.43) & 3.81 (3.64-4.00) \\
D10& 245.4, 0.84 & 1.22 (1.21-1.24) & 2.10 (2.07-2.12) & 1.21 (1.19-1.24) & 0.29 (0.28-0.29) & 0.04 (0.04-0.04) & 2.87 (2.54-3.32) &40 (4.57-500) \\
E1 & 579.9, 1.67 & 1.18 (1.16-1.20) & 1.28 (1.26-1.31) & 3.40 (3.10-3.69) & 0.39 (0.37-0.40) & 0.07 (0.06-0.08) & 5.42 (5.07-5.51) & 1.34 (1.32-1.39) \\
E2 & 401.9, 1.31 & 1.33 (1.30-1.34) & 1.21 (1.13-1.26) & 5.67 (5.34-12.86) & 0.61 (0.59-0.66) & 0.24 (0.20-0.27) & 3.21 (3.14-3.29) & 1.93 (1.89-2.03) \\
E3 & 253.6, 0.96 & 1.10 (1.05-1.16) & 1.28 (1.23-1.35) & 5.00 (3.51-6.85) & 0.48 (0.40-0.53) & 0.10 (0.09-0.15) & 5.22 (4.29-5.51) & 1.21 (1.19-1.35) \\
E4 & 192.9, 0.77 & 1.15 (1.12-1.20) & 1.27 (1.23-1.29) & 5.39 (4.54-5.68) & 0.58 (0.54-0.65) & 0.17 (0.16-0.22) & 5.37 (4.36-5.53) & 1.12 (1.10-1.16) \\
E5 & 174.8, 0.76 & 1.63 (1.61-1.64) & 1.51 (1.47-1.52) & 2.81 (2.69-2.91) & 1.21 (1.18-1.40) & 1.88 (1.77-2.49) & 0.63 (0.61-0.67) &35.38 (1.47-500) \\
E6 & 504.3, 1.51 & 1.25 (1.23-1.28) & 1.22 (1.19-1.25) & 3.39 (3.23-3.52) & 0.40 (0.40-0.41) & 0.10 (0.09-0.10) & 3.35 (3.20-3.62) & 2.44 (2.32-2.50) \\
E7 & 941.0, 2.54 & 1.13 (1.12-1.14) & 1.28 (1.28-1.32) & 2.24 (2.21-2.41) & 0.32 (0.31-0.33) & 0.06 (0.05-0.06) & 3.21 (3.14-3.26) & 3.54 (3.42-3.66) \\
F1 & 477.4, 1.63 & 1.07 (1.06-1.08) & 1.02 (1.01-1.02) & 4.96 (4.65-5.13) & 0.42 (0.41-0.43) & 0.05 (0.05-0.06) & 2.47 (2.40-2.54) & 5.77 (5.21-6.83) \\
F2 & 306.9, 1.25 & 1.09 (1.08-1.10) & 0.89 (0.88-0.89) &11.47 (10.18-13.27) & 0.58 (0.55-0.59) & 0.09 (0.08-0.09) & 2.28 (2.17-2.36) & 7.24 (6.01-12.29) \\
F3 & 191.6, 0.84 & 0.99 (0.98-1.03) & 1.02 (1.01-1.05) & 3.25 (3.00-3.40) & 0.49 (0.43-0.60) & 0.09 (0.07-0.09) & 2.48 (2.30-2.71) &43.54 (5.89-500) \\
F4 & 129.0, 0.61 & 1.03 (1.00-1.04) & 1.05 (1.04-1.06) & 3.57 (2.56-3.76) & 0.52 (0.51-0.54) & 0.10 (0.09-0.10) & 1.87 (1.83-1.99) & 7.60 (6.18-500) \\
F5 & 186.4, 0.82 & 1.01 (0.99-1.04) & 1.14 (1.10-1.16) & 2.64 (2.47-2.80) & 0.31 (0.28-0.33) & 0.07 (0.06-0.07) & 3.11 (2.52-3.69) & 2.31 (1.84-5.14) \\
G1 & 430.6, 1.66 & 1.49 (1.48-1.50) & 1.28 (1.27-1.29) & 0.70 (0.69-0.71) & 0.44 (0.43-0.45) & 0.23 (0.22-0.23) & 1.88 (1.80-2.00) &68.41 (9.04-500) \\
G2 & 244.3, 1.08 & 1.35 (1.34-1.36) & 1.20 (1.14-1.27) & 1.56 (1.48-1.62) & 0.39 (0.36-0.40) & 0.18 (0.15-0.18) & 1.22 (1.15-1.28) & 0.23 (0.19-0.27) \\
G3 & 162.2, 0.71 & 0.93 (0.92-0.94) & 1.02 (0.95-1.06) & 2.78 (2.61-2.95) & 0.48 (0.47-0.52) & 0.07 (0.06-0.07) &10.86 (10.84-12.74) & 0.43 (0.40-0.45) \\
G4 & 295.1, 1.31 & 1.03 (1.02-1.05) & 0.75 (0.75-0.76) &17.92 (13.49-34.41) & 0.70 (0.54-0.71) & 0.09 (0.08-0.09) & 2.26 (2.07-2.37) &68.41 (7.48-500) \\
G5 & 460.8, 1.68 & 1.50 (1.49-1.51) & 1.59 (1.56-1.60) & 0.51 (0.51-0.52) & 0.44 (0.43-0.44) & 0.22 (0.21-0.22) & 2.06 (2.00-2.13) & 3.35 (3.14-3.56) \\
H1 & 340.9, 0.97 & 1.91 (1.90-1.92) & 2.15 (2.12-2.23) & 0.62 (0.61-0.64) & 0.08 (0.08-0.09) & 0.04 (0.04-0.04) & 0.59 (0.58-0.59) &75.18 (26.20-500) \\
H2 & 287.0, 1.07 & 1.75 (1.74-1.77) & 1.53 (1.48-1.59) & 0.42 (0.41-0.43) & 0.40 (0.39-0.41) & 0.17 (0.16-0.17) & 1.58 (1.41-1.77) &50.18 (5.75-500) \\
H3 & 339.4, 1.29 & 1.70 (1.69-1.72) & 1.48 (1.45-1.50) & 0.47 (0.46-0.47) & 0.48 (0.48-0.49) & 0.14 (0.13-0.14) & 1.68 (1.59-1.78) &50.05 (8.09-500) \\
H4 & 333.1, 1.21 & 1.70 (1.69-1.71) & 1.55 (1.52-1.59) & 0.57 (0.55-0.58) & 0.22 (0.22-0.22) & 0.12 (0.12-0.12) & 2.76 (2.61-2.92) & 1.57 (1.46-1.67) \\
H5 & 532.9, 1.93 & 1.53 (1.52-1.55) & 0.99 (0.98-1.00) & 1.16 (1.13-1.18) & 0.20 (0.20-0.21) & 0.09 (0.09-0.09) & 2.73 (2.63-2.84) & 1.63 (1.55-1.72) \\
H6 & 323.6, 1.16 & 1.45 (1.40-1.48) & 1.12 (1.07-1.16) & 1.85 (1.68-2.12) & 0.17 (0.17-0.17) & 0.06 (0.04-0.08) & 1.36 (1.32-1.41) & 0.22 (0.20-0.23) \\
J1 & 438.1, 1.02 & 1.83 (1.82-1.86) & 2.57 (2.49-2.61) & 0.79 (0.77-0.80) & 0.10 (0.09-0.10) & 0.01 (0.01-0.01) & 3.57 (2.72-3.93) & 1.95 (1.67-2.90) \\
J2 & 537.0, 1.25 & 2.83 (2.82-2.85) & 2.31 (2.25-2.33) & 1.14 (1.13-1.18) & 0.21 (0.20-0.22) & 0.11 (0.09-0.11) & 1.86 (1.79-1.93) &40 (7.53-500) \\
J3 & 499.2, 1.13 & 1.91 (1.90-1.96) & 2.64 (2.61-2.66) & 0.95 (0.93-0.96) & 0.10 (0.10-0.10) & 0.00 (0.00-0.00) & 3.59 (3.32-4.04) & 5.20 (4.49-500) \\
J4 & 347.9, 0.90 & 1.71 (1.68-1.73) & 1.85 (1.84-1.87) & 1.52 (1.50-1.55) & 0.14 (0.14-0.14) & 0.00 (0.00-0.01) &10.85 (9.73-15.12) & 1.63 (1.43-1.91) \\
J5 & 440.6, 1.16 & 1.73 (1.68-1.79) & 1.66 (1.63-1.72) & 1.85 (1.77-1.93) & 0.13 (0.13-0.14) & 0.00 (0.00-0.00) &10.68 (10.65-10.77) & 0.71 (0.68-0.75) \\
J6 & 550.2, 1.47 & 3.24 (3.23-3.25) & 1.12 (1.12-1.13) & 4.85 (4.70-5.09) & 2.68 (2.65-2.72) & 3.90 (3.84-8.26) &10.85 (10.82-12.92) & 0.53 (0.52-0.56) \\
J7 & 756.6, 2.04 & 3.23 (3.22-3.25) & 1.14 (1.13-1.15) & 2.30 (2.25-2.33) & 0.72 (0.71-0.83) & 1.26 (1.23-1.28) & 4.29 (4.06-4.46) & 1.08 (1.04-1.12) \\
J8 & 654.7, 1.69 & 3.32 (3.30-3.33) & 1.07 (1.07-1.08) & 6.22 (6.01-6.45) & 1.61 (1.60-1.64) & 1.74 (1.70-1.78) &10.86 (10.83-11.87) & 0.51 (0.50-0.52) \\
\enddata
\end{deluxetable}

\setcounter{table}{0}
\hoffset=-0.85in
\begin{deluxetable}{cccccccccc}
\tabletypesize{\tiny}
\tablecaption{continued}
\tablewidth{8true in}
\tablehead{
\colhead{Knot}&\colhead{$\chi^2$}&\colhead{$N_H$}&\colhead{$kT$}&\colhead{$n_et$}&\colhead{Si}&\colhead{Fe} &\colhead{$kT_{Fe}$}&\colhead{$n_et_{Fe}$} \\ 
\colhead{}&\colhead{}&\colhead{(cm$^{-2}$)}&\colhead{(keV)}&\colhead{(cm$^{-3}$s)}&\colhead{(rel $\odot$)}&\colhead{(rel $\odot$)} &\colhead{(keV)}&\colhead{(cm$^{-3}$s)} \\ }
\startdata
K1 & 281.8, 0.75 & 1.97 (1.94-1.99) & 0.82 (0.81-0.83) &96.01 (18.49-500) & 0.09 (0.09-0.10) & 0.09 (0.07-0.09) &10.85 (10.85-10.90) & 0.03 (0.02-0.03) \\
K2 & 204.2, 0.61 & 1.54 (1.50-1.56) & 1.43 (1.37-1.45) & 1.51 (1.44-1.57) & 0.06 (0.06-0.06) & 0.02 (0.02-0.02) &20.61 (20.09-21.08) & 0.47 (0.45-0.49) \\
K3 & 288.5, 0.95 & 2.06 (2.05-2.08) & 1.24 (1.21-1.26) & 1.84 (1.76-1.96) & 0.10 (0.09-0.10) & 0.06 (0.05-0.06) & 1.37 (1.33-1.39) & 0.17 (0.14-0.20) \\
K4 & 300.3, 0.92 & 1.53 (1.52-1.54) & 1.48 (1.46-1.50) & 1.59 (1.55-1.62) & 0.09 (0.09-0.09) & 0.00 (0.00-0.00) & 4.44 (4.11-4.67) & 1.50 (1.41-1.56) \\
K5 & 332.6, 1.08 & 1.59 (1.55-1.60) & 1.27 (1.24-1.29) & 1.51 (1.45-1.56) & 0.08 (0.08-0.09) & 0.01 (0.01-0.02) & 3.17 (3.00-3.40) & 2.23 (2.08-3.28) \\
L1 & 241.2, 0.79 & 1.63 (1.62-1.65) & 1.62 (1.57-1.65) & 1.03 (1.00-1.09) & 0.04 (0.03-0.04) & 0.02 (0.02-0.02) & 0.14 (0.14-0.16) & 0.26 (0.26-0.26) \\
L2 & 301.5, 1.05 & 1.54 (1.53-1.55) & 1.39 (1.36-1.41) & 0.70 (0.69-0.72) & 0.07 (0.06-0.07) & 0.08 (0.08-0.09) & 2.77 (1.57-2.95) & 2.08 (1.94-2.24) \\
L3 & 584.6, 1.78 & 1.46 (1.45-1.47) & 2.00 (1.98-2.03) & 0.55 (0.55-0.56) & 0.14 (0.14-0.14) & 0.08 (0.08-0.08) & 2.75 (2.68-2.85) & 2.25 (2.14-3.68) \\
L4 & 412.4, 1.31 & 1.55 (1.54-1.56) & 2.55 (2.47-2.61) & 0.48 (0.47-0.49) & 0.20 (0.19-0.20) & 0.08 (0.07-0.08) & 1.88 (1.77-1.95) &40 (7.95-500) \\ 
L5 & 424.9, 1.18 & 1.60 (1.59-1.61) & 2.73 (2.66-2.79) & 0.50 (0.49-0.51) & 0.10 (0.10-0.10) & 0.05 (0.05-0.05) & 2.18 (2.12-2.24) & 3.98 (3.63-4.34) \\
L6 & 400.6, 1.21 & 1.49 (1.48-1.50) & 2.53 (2.50-2.58) & 0.60 (0.59-0.61) & 0.18 (0.17-0.18) & 0.09 (0.09-0.09) & 4.31 (4.04-4.57) & 1.41 (1.34-1.47) \\
M1 & 495.3, 1.51 & 1.32 (1.31-1.33) & 2.75 (2.68-2.79) & 0.61 (0.61-0.62) & 0.20 (0.20-0.21) & 0.08 (0.08-0.08) & 2.50 (2.41-2.57) & 3.70 (3.47-3.95) \\
M2 & 543.3, 1.81 & 1.43 (1.42-1.44) & 2.73 (2.69-2.77) & 0.44 (0.43-0.44) & 0.30 (0.30-0.31) & 0.09 (0.09-0.10) & 2.04 (1.98-2.10) & 6.65 (5.84-8.14) \\
M3 & 409.8, 1.35 & 1.42 (1.41-1.43) & 2.88 (2.80-2.92) & 0.46 (0.45-0.46) & 0.25 (0.24-0.25) & 0.08 (0.07-0.08) & 2.19 (2.12-2.30) & 3.92 (3.49-6.08) \\
M4 & 373.7, 1.33 & 1.34 (1.33-1.35) & 2.92 (2.87-3.00) & 0.48 (0.47-0.49) & 0.35 (0.34-0.36) & 0.12 (0.11-0.12) & 2.37 (2.27-2.50) & 3.66 (3.41-4.04) \\
M5 & 387.4, 1.40 & 1.29 (1.29-1.31) & 2.57 (2.52-2.63) & 0.52 (0.52-0.53) & 0.36 (0.36-0.37) & 0.13 (0.12-0.13) & 2.18 (2.08-2.32) &56.02 (6.47-500) \\
M6 & 381.8, 1.34 & 1.21 (1.21-1.23) & 2.15 (2.10-2.17) & 0.84 (0.77-0.87) & 0.34 (0.33-0.35) & 0.08 (0.07-0.08) & 0.61 (0.60-0.62) &60 (2.41-500) \\
M7 & 182.7, 0.69 & 1.09 (1.04-1.13) & 1.54 (1.46-1.66) & 1.99 (1.72-2.09) & 0.14 (0.13-0.14) & 0.04 (0.04-0.04) & 1.92 (1.40-2.09) & 0.27 (0.26-0.30) \\
N1 & 807.4, 2.21 & 1.25 (1.24-1.25) & 2.41 (2.38-2.43) & 0.55 (0.54-0.55) & 0.22 (0.21-0.22) & 0.07 (0.07-0.08) & 1.93 (1.91-1.98) &68.41 (8.95-500) \\
N2 & 670.3, 2.26 & 1.42 (1.41-1.42) & 2.16 (2.15-2.19) & 0.36 (0.36-0.37) & 0.46 (0.45-0.46) & 0.08 (0.07-0.08) & 1.78 (1.72-1.86) &34.84 (10.38-500) \\
N3 & 483.4, 1.67 & 1.43 (1.43-1.44) & 2.15 (2.12-2.16) & 0.49 (0.49-0.50) & 0.42 (0.42-0.43) & 0.09 (0.08-0.09) & 2.34 (2.23-2.57) & 3.62 (3.31-4.59) \\
N4 & 389.4, 1.36 & 1.20 (1.20-1.21) & 2.01 (1.98-2.04) & 0.66 (0.66-0.67) & 0.38 (0.37-0.38) & 0.08 (0.08-0.08) & 2.10 (1.98-2.21) &30 (6.75-500) \\
N5 & 359.4, 1.40 & 1.24 (1.23-1.26) & 1.81 (1.78-1.84) & 0.66 (0.64-0.67) & 0.50 (0.49-0.51) & 0.06 (0.05-0.06) & 2.10 (1.82-2.28) &49.18 (6.30-500) \\
\enddata
\end{deluxetable}

\end{document}